\documentclass[aps,prd, 
superscriptaddress, 
showpacs,
preprintnumbers,
nofootinbib,
bibnotes,
amsmath,
amssymb,
floatfix
]{revtex4-2}

\usepackage[utf8]{inputenc}
\usepackage{array,bbold,booktabs}
\usepackage{graphicx}
\usepackage{dcolumn}
\usepackage{bm}
\usepackage{longtable}
\usepackage{natbib}
\usepackage{hyperref}
\usepackage[capitalise]{cleveref}
\crefname{section}{Sec.}{Secs.}

\usepackage{todonotes} 

\newcommand{\amuSD}{a_\mu^{\text{SD}}}
\newcommand{\amuSDq}{a_\mu^{\text{SD,q}}}

\newcommand{\amuSDqblinded}{a_\mu^{\text{SD,q,blinded}}}

\newcommand\regensburg{Department of Physics, University of Regensburg, 93040 Regensburg, Germany}

\begin{document}
\title{A high-precision continuum limit study of the HVP short-distance window}

\author{Sebastian Spiegel}\email{sebastian.spiegel@ur.de}\affiliation{\regensburg}
\author{Christoph Lehner}\email{christoph.lehner@ur.de}\affiliation{\regensburg}

\date{\today}

\begin{abstract}
The separation of the hadronic vacuum polarization (HVP) contribution to the muon anomalous magnetic moment into Euclidean windows allows for a tailored approach to address the different dominant challenges at short, intermediate, and long distances.  We present a novel approach to compute the short-distance window  without the need for using perturbative QCD.  We combine a quenched continuum extrapolation using 18 lattice spacings ($1.6 \,\text{GeV} \lesssim a^{-1} \lesssim 6.1\,\text{GeV}$) with a separate continuum extrapolation of the sea-quark effects.  This method allows for the computationally expensive sea-quark effects to be estimated using only a smaller number of ensembles at coarser lattice spacings, while largely confining the logarithmic dependency of the continuum extrapolation to the quenched component.
\end{abstract}

\maketitle

\section{Introduction} \label{sec:Introduction}
The magnetic moment of the muon is defined by
\begin{align}
\bm{\mu} = - g_\mu \frac{e}{2 m_\mu} \bm{S}\,,
\end{align}
where $e$ is the electric charge, $m_\mu$ is the muon mass, $\bm{S}$ is the particle’s spin, and $g_\mu$ denotes its Landé factor.  The deviation with Dirac's result of $g_\mu=2$ is defined as the anomalous magnetic moment of the muon,
\begin{equation}
    a_\mu = \frac{g_\mu - 2}{2}\,,
\end{equation}
which captures the radiative corrections of all known and unknown particles. This quantity is one of the most precisely determined values in physics and plays an important role in high-precision
tests of our understanding of physics at the fundamental level.  
The latest experimental results provided by the FNAL Muon g-2 Collaboration, which combine their first two publications \cite{PhysRevLett.126.141801,PhysRevLett.131.161802} with the final result of the BNL E821 experiment \cite{PhysRevD.73.072003}, arrive at a precision of $0.19$ parts per million (ppm) for the world average.

In contrast, the recommended Standard Model result of the Muon g-2 Theory Initiative from 2020 \cite{Aoyama2020} has an uncertainty of 0.37 ppm. To match the precision of the experimental measurement, the theoretical uncertainty must be reduced by roughly half. This, combined with the anticipated progress from the planned J-PARC experiment \cite{abe2019newapproachmeasuringmuon}, underscores the need for novel methods to improve the theoretical uncertainty. The leading-order hadronic vacuum polarization (HVP) predominantly contributes to the theoretical prediction’s uncertainty, making its refinement particularly urgent. \\
Using the Euclidean window approach introduced by the RBC/UKQCD collaborations in \cite{blum2018hvp}, this quantity can be divided into three sub-contributions: short-distance, window, and long-distance contributions. This allows to separate the challenging short- and long-distance quantities from the window quantity which has tractable statistical and systematic errors. While the discretization effects are most significant for the short-distance contribution, the statistical noise and finite-volume effects are particularly severe in the long-distance contribution. This work aims to scrutinize the continuum limit of the short-distance window from first principles using purely Lattice QCD methods, without perturbative input. The approach combines a precise quenched continuum limit, derived from 18 lattice spacings, with a separate continuum extrapolation of the dynamical sea quark corrections, utilizing only a few computationally expensive ensembles. \\
This paper is organized as follows: In \cref{sec:Methodology}, we introduce the methodology of the time-momentum representation for the leading-order HVP, which facilitates the definition of Euclidean windows. The computational details are provided in \cref{sec:ComputationalDetails}. Subsequently, we present the blinded quenched and sea quark continuum limit results, along with our blinded short-distance HVP results, before unblinding the latter.

\section{Methodology}\label{sec:Methodology}
In the following sections, we describe the time-momentum representation (\cref{sec:TimeMomentumRepresentation}) and define the Euclidean windows (\cref{sec:EuclideanWindows}), which provide a basis for computing the short-distance contribution to the leading-order HVP $a_\mu^\text{HVP LO}$. For brevity, we will omit the superscript HVP LO from this point onward.

\subsection{Time-momentum representation}\label{sec:TimeMomentumRepresentation}
We consider the correlator
\begin{align}
    C(t) = \frac{1}{3} \sum_{\vec{x}} \sum_{j=0,1,2} \langle J_j(\vec{x}, t) J_j(0) \rangle,
\end{align}
where $J_\mu(x)$ denotes the vector current, defined as 
\begin{align}
    J_\mu(x) = i \sum_f Q_f \bar{\psi}_f(x) \gamma_\mu \psi_f(x).
    \label{eq:VectorCurrent}
\end{align}
In this expression, $Q_f$ denotes the fractional electric charge, and the sum runs over the quark flavors $f$. The leading-order HVP contribution to $a_\mu$ is then computed as
\begin{align}
    a_\mu = \sum_{t=0}^\infty w_t C(t).
    \label{eq:a_mu}
\end{align}
The weights $w_t$ incorporate the photon and muon effects within the HVP diagrams and can be computed via the integral \cite{bernecker2011vectorcorrelators}
\begin{equation}
    w_t = 8 \alpha^2 \int_0^\infty \textbf{d}Q^2 \left(\frac{\cos(Q t) - 1}{Q^2} + \frac{1}{2} t^2\right) f(Q),
\end{equation}
where $Q^2$ denotes the Euclidean photon four-momentum squared, $\alpha$ is the fine structure constant, and 
\begin{align}
    f(Q) = \frac{m_\mu^2 Q^2 Z^3(Q) \left( 1 - Q^2 Z(Q) \right)}{1 + m_\mu^2 Q^2 Z^2(Q)} \quad \text{with} \quad Z(Q) = \frac{\sqrt{Q^4 + 4Q^2m_\mu^2} - Q^2}{2m_\mu^2 Q^2}.
\end{align}
Here, $m_\mu$ abbreviates the muon mass. Additionally, we consider an alternative weight definition:
\begin{align}
    \hat{w}_t = 8 \alpha^2 \int_{0}^{\infty} \text{d}Q^2 \left( \frac{\cos(Qt) - 1}{(2 \sin(Q/2))^2} + \frac{1}{2}t^2 \right) f(Q)
    \label{eq:w_hat}
\end{align}
which uses a lattice discretization of the photon momentum and yields the same value of $a_\mu$ in the continuum limit. Using both weight definitions allows us to investigate the continuum limit and test the robustness of our analysis approach. \\
In this study, we compute the up and down quark-connected contribution to the correlator $C(t)$ in the isospin-symmetric limit. For brevity, we omit an explicit label in the notation.

\subsection{Euclidean windows}\label{sec:EuclideanWindows}
In line with the method outlined in \cite{blum2018hvp}, we utilize Euclidean windows to separate the contributions from time slices t in \cref{eq:a_mu} into short-distance (SD), window (W), and long-distance (LD) components. We apply smearing kernels with a width $\Delta$ to ensure well-defined quantities at non-zero lattice spacing. This approach leads to the partitioning:
\begin{align}
    a_\mu = a_\mu^\text{SD} + a_\mu^\text{W} + a_\mu^\text{LD}
\end{align}
with the individually well-defined contributions:
\begin{align}
    a_\mu^\text{SD}(t_0,\Delta) &= \sum_{t=0}^\infty C(t) w_t [1 - \Theta(t,t_0,\Delta)], \\
    a_\mu^\text{W}(t_0,t_1,\Delta) &= \sum_{t=0}^\infty C(t) w_t [\Theta(t,t_0,\Delta) - \Theta(t,t_1,\Delta)], \\
    a_\mu^\text{LD}(t_1,\Delta) &= \sum_{t=0}^\infty C(t) w_t \Theta(t,t_1,\Delta), \\ 
    \Theta(t,t',\Delta) &= [1 + \tanh[(t-t')/\Delta]]/2.
\end{align}
This work focuses on determining the short-distance segment $a_\mu^\text{SD}(t_0, \Delta)$, relying exclusively on lattice methods. Among the contributions, discretization effects are expected to be most severe for this segment. For our analysis, we use the values $t_0 = 0.4\ \text{fm}$ and $\Delta = 0.15\ \text{fm}$ \cite{blum2018hvp}.

\section{Computational details}\label{sec:ComputationalDetails}
In the following, the computational details of this work are discussed. This includes ensemble generation, an overview of the analyzed measurements, as well as crucial aspects of the analysis.

\subsection{Data description}\label{sec:DataDescription}
The quenched ensembles for this study were generated using the Hybrid Monte Carlo (HMC) algorithm \cite{duane1987hybridmontecarlo} and the Iwasaki gauge action. Each molecular dynamics (MD) trajectory was evolved for a time of $\tau = 1.0$ using 12 steps of an 4th order Omelyan integrator \cite{omelyan2003272}. A total of 21 quenched ensembles with lattice spacings ranging from $a^{-1} \approx 1.57$ to $6.10\,\text{GeV}$ were generated, with an overview provided in \cref{tab:QuenchedData}. Among these ensembles, 9 have periodic boundary conditions (periodic BC) and a lattice volume of $24^3 \times 48$. The inverse couplings for these ensembles range from $\beta = 2.40$ to $\beta = 2.80$ in steps of $0.05$. Additionally, 8 ensembles with open boundary conditions (open BC) and a lattice volume of $24^3 \times 96$ were generated, with $\beta$ values ranging from $2.80$ to $3.15$, also in steps of $0.05$. To further refine our analysis, 3 additional open BC ensembles with inverse couplings $\beta = 3.0, 3.15,$ and $3.2$ were generated using a larger lattice volume of $32^3 \times 96$. Finally, an ensemble with a lattice volume of $48^3 \times 192$ at $\beta = 3.4$ was generated, also with open BC. \\
To decrease autocorrelations, only every 100th generated MD trajectory was considered for subsequent analyses of each ensemble, except for $\beta = 3.4$ where only every 250th was retained. Thermalization of the Markov chains was decided based on gradient flow scale measurements (cf. \cref{sec:ScaleSetting}) on each configuration. For open BC ensembles, all measurements were performed within their respective bulk, which we define by the time slices $[\frac{T}{2a} - \frac{L}{a}, \frac{T}{2a} + \frac{L}{a}]$, unless indicated otherwise. \\ 
The analysis presented consists of two main types of measurements: purely gluonic measurements based on the energy density and vector correlator measurements. For the gluonic measurements, every second available configuration was used to further reduce correlations. The fermionic measurements were performed on consecutive configurations. The last two columns of \cref{tab:QuenchedData} show the sample sizes for both gluonic and fermionic measurements. For the fermionic measurements, we employed Möbius \cite{brower2014mobiusdomainwallfermion} domain-wall \cite{shamir1993chiralfermions,furman1995axialsymmetries} fermions with parameters $b=1.5$ and $c=0.5$. The length of the fifth dimension, which controls residual chiral-symmetry-breaking effects, was set to $L_s=12$ for all quenched ensembles. We verified that the residual chiral-symmetry-breaking effects are small even for the finest lattice ensemble by evaluating the correlators at $\beta=3.4$ also with $L_s=24$.  All observed variations were sub-per-mille. \\
In our analysis, we use four dynamical ensembles that were generated using the Iwasaki gauge action and Möbius domain-wall fermion sea quarks. These ensembles employ $N_f = 2 + 1$ sea quark flavors, with lattice spacings ranging from $a^{-1} \approx 1.73$ to $3.53\,\text{GeV}$. A detailed description of ensemble 4 can be found in \cite{blum2023update}, while the other ensembles were produced with similar codes. \cref{tab:DynamicalData} provides an overview of the dynamical ensembles, along with their properties.

\begin{table}
{\setlength{\tabcolsep}{1em}
  \begin{tabular}{l|llllllll}
  \midrule\midrule
  $\beta$ & $a^{-1}$/GeV & $\sqrt{t_0}/a$ & $w_0/a$ & $L^3 \times T/a^4$ & BC & $N_\text{gluonic}$ & $N_\text{fermionic}$ \\\hline
  $2.40$ & $1.5677(95)$ & $1.30154(53)$ & $1.28618(72)$ & $24^3 \times 48$ & periodic & $200$ & $100$ \\
  $2.45$ & $1.691(11)$ & $1.40422(80)$ & $1.3931(12)$ & $24^3 \times 48$ & periodic & $200$ & $100$ \\
  $2.50$ & $1.824(12)$ & $1.5139(10)$ & $1.5068(15)$ & $24^3 \times 48$ & periodic & $200$ & $100$ \\
  $2.55$ & $1.966(13)$ & $1.6324(11)$ & $1.6313(16)$ & $24^3 \times 48$ & periodic & $200$ & $100$ \\
  $2.60$ & $2.109(13)$ & $1.7507(16)$ & $1.7523(24)$ & $24^3 \times 48$ & periodic & $200$ & $100$ \\
  $2.65$ & $2.265(14)$ & $1.8807(20)$ & $1.8881(30)$ & $24^3 \times 48$ & periodic & $200$ & $200$ \\
  $2.70$ & $2.425(16)$ & $2.0136(26)$ & $2.0239(36)$ & $24^3 \times 48$ & periodic & $200$ & $100$ \\
  $2.75$ & $2.601(17)$ & $2.1595(34)$ & $2.1764(50)$ & $24^3 \times 48$ & periodic & $200$ & $100$ \\
  $2.80$ & $2.781(18)$ & $2.3091(36)$ & $2.3303(53)$ & $24^3 \times 48$ & periodic & $420$ & $220$ \\\hline
  $2.80$ & $2.780(18)$ & $2.3083(40)$ & $2.3254(59)$ & $24^3 \times 96$ & open & $200$ & $230$ \\
  $2.85$ & $2.969(20)$ & $2.4652(58)$ & $2.4835(82)$ & $24^3 \times 96$ & open & $200$ & $200$ \\
  $2.90$ & $3.184(22)$ & $2.6434(71)$ & $2.6739(91)$ & $24^3 \times 96$ & open & $200$ & $200$ \\
  $2.95$ & $3.394(24)$ & $2.8179(90)$ & $2.849(13)$ & $24^3 \times 96$ & open & $200$ & $200$ \\
  $3.00$ & $3.650(26)$ & $3.0303(97)$ & $3.077(15)$ & $24^3 \times 96$ & open & $200$ & $200$ \\
  $3.05$ & $3.901(30)$ & $3.238(15)$ & $3.289(22)$ & $24^3 \times 96$ & open & $200$ & $200$ \\
  $3.10$ & $4.142(32)$ & $3.439(16)$ & $3.492(23)$ & $24^3 \times 96$ & open & $200$ & $400$ \\
  $3.15$ & $4.462(43)^*$ & $3.704(28)^*$ & $3.780(42)^*$ & $24^3 \times 96$ & open & $200$ & $200$ \\\hline
  $3.00$ & $3.635(24)$ & $3.0174(74)$ & $3.055(11)$ & $32^3 \times 96$ & open & $200$ & $200$ \\
  $3.15$ & $4.433(31)$ & $3.680(13)$ & $3.74(18)$ & $32^3 \times 96$ & open & $200$ & $300$ \\
  $3.20$ & $4.737(39)^*$ & $3.932(21)^*$ & $4.004(30)^*$ & $32^3 \times 96$ & open & $200$ & $400$ \\\hline
  $3.40$ & $6.099(43)^*$ & $5.063(17)^*$ & $5.139(25)^*$ & $48^3 \times 192$ & open & $200$ & $200$ \\
  \midrule\midrule
  \end{tabular}}
\caption{List of quenched ensembles with simulation parameters. The ensembles were generated using the Iwasaki gauge action, and their lattice spacing was determined employing the gradient flow scale $t_0$ with physical input from \cite{sommer2014scalesettinglatticeqcd,bruno2014nfdependencegluonicobservables,luescher2010wilsonflow}. The columns represent the inverse coupling ($\beta$), the inverse lattice spacing ($a^{-1}/\text{GeV}$), the gradient flow scales ($\sqrt{t_0}/a$ and $w_0/a$), the lattice volume ($L^3 \times T/a^4$), the boundary conditions (BC), and the sample sizes for gluonic ($N_\text{gluonic}$) and fermionic ($N_\text{fermionic}$) measurements. The asterisk ($^*$) indicates that in these cases the binning study did not conclusively converge for the scale setting parameters themselves.  We address this issue by using the delayed binning strategy described in the following.}
\label{tab:QuenchedData}
\end{table}

\begin{table}
{\setlength{\tabcolsep}{1em}
  \begin{tabular}{l|llllllll}
  \midrule\midrule
  ID & $a^{-1}$/GeV & $\sqrt{t_0}/a$ & $w_0/a$ & $N_f$ & $L^3 \times T \times L_s/a^4$ & BC & $M_\pi$/MeV & $M_K$/MeV  \\\hline
  4 & $1.7312(28)$ & $1.2880(09)$ & $1.4758(20)$ & 2+1 & $24^3 \times 48 \times 24$ & periodic & $274.8(2.5)$ & $530.1(3.1)$ \\
  9 & $2.3549(49)$ & $1.7335(09)$ & $2.0206(20)$ & 2+1 & $32^3 \times 64 \times 12$ & periodic & $278.79(62)$ & $530.98(75)$ \\
  F & $2.6920(67)$ & $1.9907(04)$ & $2.3583(08)$ & 2+1 & $48^3 \times 96 \times 12$ & periodic & $283.2(1.0)$ & $519.3(1.4)$ \\
  E & $3.53(01)$  & $2.5678(28)$ & $3.0254(76)$ & 2+1 & $48^3 \times 192 \times 12$ & open & $289.5(2.1)$ & $540.0(2.5)$ \\
  \midrule\midrule
  \end{tabular}}
\caption{List of dynamical ensembles with simulation parameters. The ensembles were generated using the Iwasaki gauge action and Möbius domain-wall fermion sea quarks. The columns represent the ensemble identifier (ID), the inverse lattice spacing ($a^{-1}$/GeV), the gradient flow scales ($\sqrt{t_0}/a$ and $w_0/a$), the number of sea quark flavors ($N_f$), the lattice dimensions ($L^3 \times T \times L_s/a^4$), the boundary conditions (BC), and the pion and kaon masses ($M_\pi$, $M_K$) in MeV.}
\label{tab:DynamicalData}
\end{table}

\subsection{Statistical and systematic errors}\label{sec:StatisticalSystematicErrors}
To obtain statistical error estimates on measured observables, we rely on a modified version of the jackknife method. For $n$ different observables $O_i$ with $i=1,....n$, we consider the sample means:
\begin{align}
    \overline{O}_i = \frac{1}{|S|} \sum_{\tau \in S} O_i^{\tau},
\end{align}
where $S$ denotes a selection of thermalized Markov times $\tau$. The covariance of two such observables $O_i$ and $O_j$ may be estimated using:
\begin{align}
    \overline{\text{Cov}}(O_i,O_j) = \frac{1}{|S|-1} \sum_{\tau \in S} \left(O_i^{\tau} - \overline{O}_i\right) \left(O_j^{\tau} - \overline{O}_j\right).
\end{align}
Consider now a function: 
\begin{align}
    \overline{f}_\alpha = f_\alpha(\overline{O}_1,...,\overline{O}_n),
\end{align}
for which we also wish to estimate the corresponding covariance. The standard jackknife estimation procedure consists of computing the single-elimination jackknife average defined according to 
\begin{align}
    \overline{O}_i^{(\tau)} = \frac{1}{|S|-1} (|S|\overline{O}_i - O_i^\tau) = \overline{O}_i + \frac{1}{|S|-1}(\overline{O}_i - O_i^\tau)
\end{align}
and correspondingly 
\begin{align}
    \overline{f}_\alpha^{(\tau)} = f_\alpha(\overline{O}_1^{(\tau)},...,\overline{O}_n^{(\tau)}).
\end{align}
The jackknife estimate for the covariance between $\overline{f}_\alpha$ and $\overline{f}_\beta$ is subsequently given by
\begin{align}
    \overline{\text{Cov}}(\overline{f}_\alpha, \overline{f}_\beta) = \frac{|S|-1}{|S|} \sum_{\tau \in S} \left(\overline{f}_\alpha^{(\tau)} - \overline{f}_\alpha\right) \left(\overline{f}_\beta^{(\tau)} - \overline{f}_\beta\right).
\end{align}
This method assumes that the Markov time selection $S$ is chosen in such a way that all data is uncorrelated. In practice, however, this is often not the case. One way to remedy this is by employing a binning procedure and studying the covariances as a function of the bin size $m$ to identify a plateau for large $m$. Incorporating the binning of the data with the jackknife average definition in a single steps allows to define the $m$-elimination jackknife average
\begin{align}
    \overline{O}_i^{(\tau_1,...,\tau_m)} = \frac{1}{|S|-m} \left(|S|\overline{O}_i  - \sum_{k=1}^m O_i^{\tau_k}\right) = \overline{O}_i + \frac{1}{|S|-m} \left(m \overline{O}_i - \sum_{k=1}^m O_i^{\tau_k}\right).
\end{align}
This can be used to compute $\overline{f}_\alpha^{(\tau_1,...,\tau_m)}$ and subsequently the jackknife estimate for the covariance. However, this procedure requires prior knowledge of the appropriate bin size to ensure data independence, as correlations between subsequent measurements vary by observable. To address this, we reconstruct the $m$-elimination jackknife sample using the formula:
\begin{align}
    \overline{f}_\alpha^{(\tau_1,...,\tau_m)} = \overline{f}_\alpha + \frac{|S|-1}{|S|-m}\left(\sum_{k=1}^{m} \overline{f}_\alpha^{(\tau_k)} - m \overline{f}_\alpha \right) + \mathcal{O}\left(\frac{1}{(|S|-1)(|S|-m)}\right).
\end{align}
This approach allows us to delay the binning in the jackknife method and compute $\overline{f}_\alpha^{(\tau_1,...,\tau_m)}$ directly from the unbinned $\overline{f}_\alpha^{(\tau)}$. Consequently, only unbinned jackknife samples need to be stored, which enables us to study autocorrelations using the reconstructed binned jackknife sample. \\
Next, we briefly discuss the treatment of systematic errors in our analysis. These errors arise from external inputs, modeling assumptions, or computational methods. Given that their values are much smaller compared to the sample means, we assume that derived quantities can be linearized for sufficiently small deviations from the mean. To propagate systematic errors to a function dependent on these means, we use the following approach for each systematic error: compute the function at the mean plus the systematic error, then compute the quadratic difference between this result and the function evaluated at the mean. This quadratic difference is treated as the variance, assuming that systematic errors are symmetrically distributed around the mean.

\subsection{Scale setting}\label{sec:ScaleSetting}
\begin{figure}
    \centering
    \includegraphics[width=.8\textwidth]{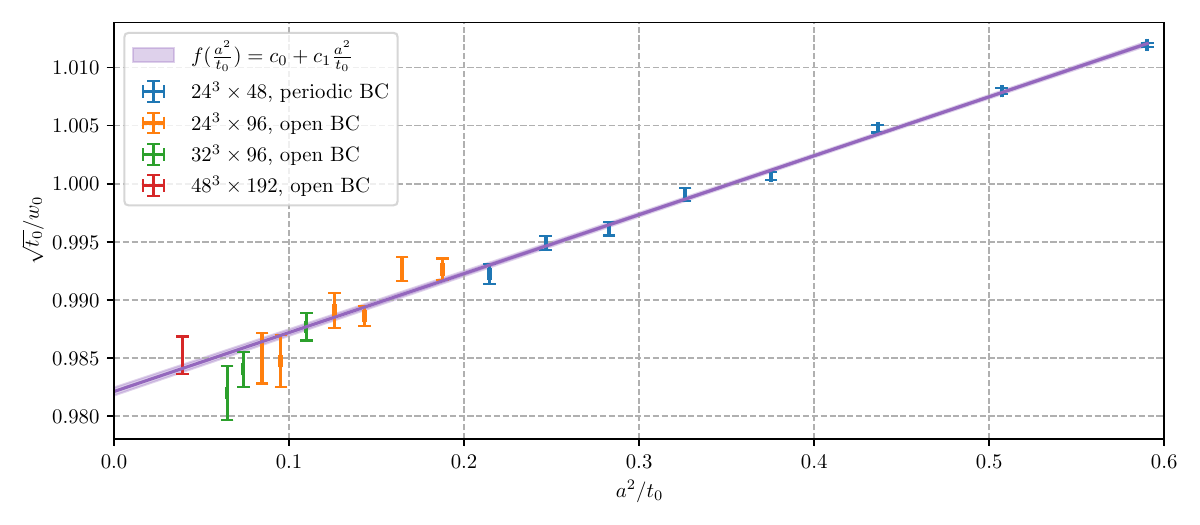}
    \caption{Continuum extrapolation of $\sqrt{t_0}/w_0$ with linear fit model $f(a^2/t_0) = c_0 + c_1 a^2/t_0$.}
    \label{fig:sqrt_t0_by_w0_fit}
\end{figure}

The scale in this analysis is set using the gradient flow scales $t_0$ and $w_0$. The scale $t_0$ is defined \cite{luescher2010wilsonflow} by the condition
\begin{align}
    t^2 \left\langle E(t) \right\rangle \big|_{t=t_0} = 0.3,
    \label{eq:t0}
\end{align}
where $E(t)$ denotes the Wilson flow smeared energy density at flow time $t$. Similarly, the scale $w_0$ is defined by the condition \cite{borsanyi2012highprecisionscalesetting}
\begin{align}
    t \frac{d}{dt} \left(t^2 \langle E(t)\rangle \right)\big|_{t=w_0^2} = 0.3. 
    \label{eq:w0}
\end{align}
We calculate the energy density $E(t)$ on the lattice using the cloverleaf definition. For ensembles with open BC, the averaging is restricted to the bulk. \\
We noted above that the thermalization of Markov chains was determined based on the gradient flow scales. For this purpose, a modified version of the scales was used, where the statistical average over the energy density was replaced by the energy density of a single configuration at each Markov time. These quantities exhibit particularly slow Markov modes compared to other observables considered in this work. \\
The third and fourth columns of \cref{tab:QuenchedData} show the determined estimates of $\sqrt{t_0}/a$ and $w_0/a$ for the quenched ensembles. For ensembles marked with an asterisk ($^*$) the binning study for the scale quantities did not convincingly converge. Specifically, these ensembles include the ensembles at $\beta = 3.15$ with a lattice volume of $24^3 \times 96$, $\beta = 3.2$ with a lattice volume of $32^3 \times 96$, and $\beta = 3.4$ with a lattice volume of $48^3 \times 192$. This observation is based on a binning analysis, where the maximum bin size for each ensemble was chosen such that the binned samples had a size of at least 25 to ensure a sufficiently small error on the error. All subsequent binning studies mentioned in this work were performed in the same manner.  However, this observation does not impact subsequent parts of this work, as the delayed binning in the jackknife can be applied separately for derived quantities depending on the flow scale.  For such derived quantities the delayed binning study did not indicate residual
autocorrelations such that the observation only limits the precision with which we can quote the scale quantities.
In \cref{fig:sqrt_t0_by_w0_fit}, we illustrate the continuum extrapolation of the dimensionless ratio $\sqrt{t_0}/w_0$ as a function of $a^2/t_0$ using a linear model
\begin{equation}
    f\left(\frac{a^2}{t_0}\right) = c_0 + c_1 \frac{a^2}{t_0}
\end{equation}
As $a^2/t_0$ approaches zero, the extrapolated value of $\sqrt{t_0}/w_0$ converges to a finite continuum limit. For the inverse coupling $\beta = 2.8$, the open BC measurements were considered, while for $\beta = 3.0$ and $\beta = 3.15$, we used the ensembles with larger lattice volumes. The data shows that the lattice effects appear to be of order $a^2$ over the full range of considered quenched ensembles. 
To determine the lattice spacing $a$, we rely on the gradient flow scale $t_0$. We use the following external input from \cite{sommer2014scalesettinglatticeqcd,bruno2014nfdependencegluonicobservables,luescher2010wilsonflow} to set the physical value of $\sqrt{t_0}$:
\begin{align}
    \sqrt{t_0} = 0.1638(10)\,\text{fm}.
\end{align}
While this choice defines how we translate dimensionful quantities such as the muon mass to dimensionless units in the quenched world, our final results are corrected to include dynamical sea quarks and are independent of this choice.  Note that one may use the central value of $\sqrt{t_0}$ without uncertainty as the definition of physical distance for our intermediate quenched world.

\begin{figure}
    \centering
    \includegraphics[width=.8\textwidth]{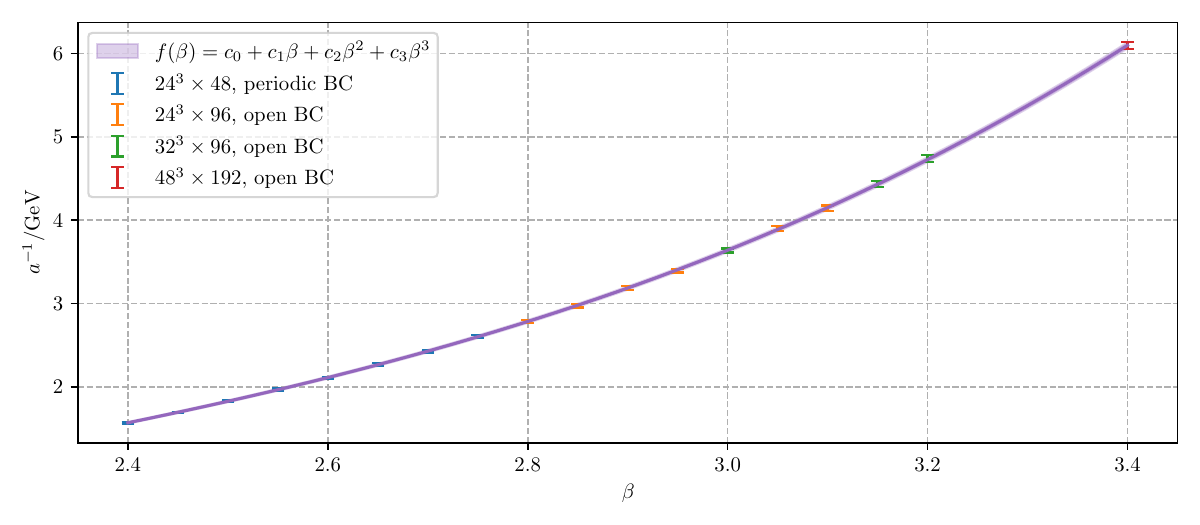}
    \caption{Interpolation of $a^{-1}/\text{GeV}$ derived from $\sqrt{t_0}$ versus $\beta$ using cubic fit model $f(\beta) = c_0 + c_1 \beta + c_2 \beta^2 + c_3 \beta^3$.}    
    \label{fig:a_inv_GeV_fit}
\end{figure}

In \cref{fig:a_inv_GeV_fit}, we present an interpolation of $a^{-1}/\text{GeV}$ derived from $\sqrt{t_0}$ versus $\beta$. The $\beta$ values range from 2.4 to 3.4. For $\beta = 2.8$, we used the open BC data, and for $\beta = 3.0$ and $\beta = 3.15$, we used the large volume ensembles. The interpolation is performed using a cubic fit model of the form:
\begin{align}
f(\beta) = c_0 + c_1 \beta + c_2 \beta^2 + c_3 \beta^3,
\end{align}
with fitted parameters $c_0 = -13.143(12)$, $c_1 = 15.8718(62)$, $c_2 = -6.5925(22)$, and $c_3 = 1.0555(18)$. The $p$-value of the fit is approximately $0.38$.

\subsection{Mass interpolation}\label{sec:MassInterpolation}
\begin{figure}
    \centering
    \includegraphics[width=.9\textwidth]{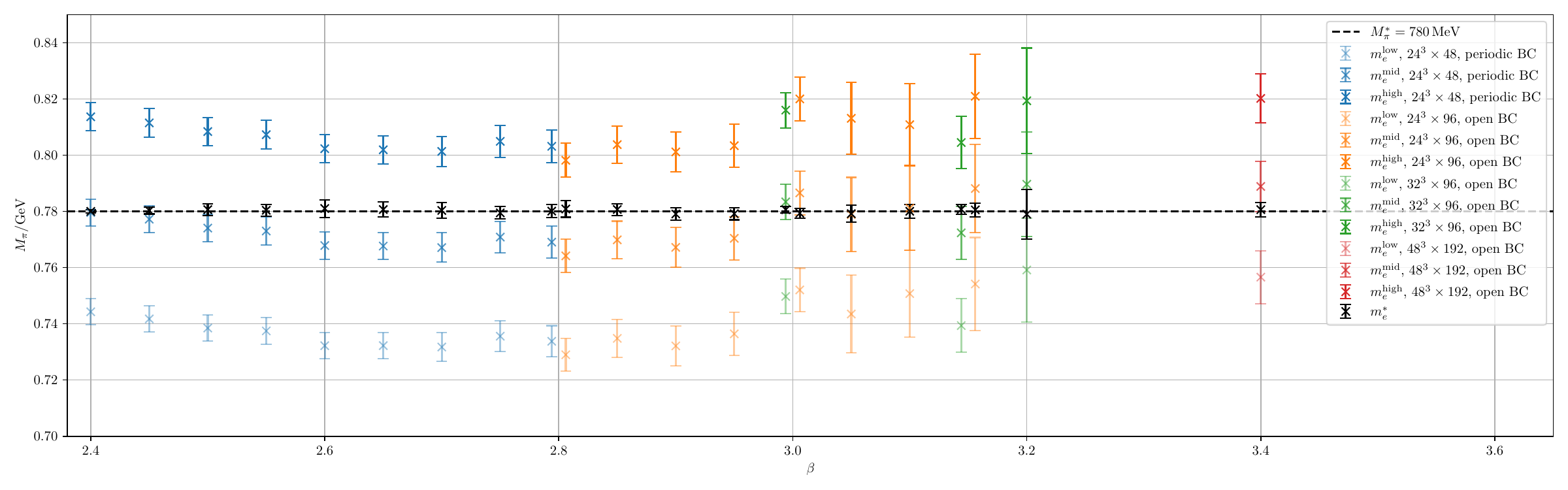}
    \caption{Pion masses for three different quark masses $m_e^\text{low}$, $m_e^\text{mid}$, and $m_e^\text{high}$ calculated for each quenched ensemble $e$. We note that for ensembles with the same inverse coupling $\beta$, the data points were shifted slightly to enhance visibility. The quark masses were selected such that a single physical pion mass of $M_\pi^* = 780\,\text{MeV}$ could be chosen to lie within the ranges defined by these three quark masses across all 21 ensembles. The plot demonstrates the consistency of the physical pion masses obtained from the quadratically interpolated pion correlators at quark masses $m_e^*$ with $M_\pi^*$.} 
    \label{fig:pion_masses}
\end{figure}

Although we expect short-distance correlators to be largely insensitive to the quark mass \cite{blum2023update}, we aimed to compute the short-distance hadronic vacuum polarization at the same physical mass point across ensembles with different lattice spacings to minimize any potential mass effects. To achieve this, we calculated pion correlators for three different quark masses $m_e^\text{low}$, $m_e^\text{mid}$ and $m_e^\text{high}$ for each quenched ensemble $e$ and determined the corresponding pion masses in physical units, see \cref{fig:pion_masses}. The quark masses were selected such that a single physical pion mass $M_\pi^*$ could be chosen to lie within the ranges defined by the three quark masses for all of the 21 ensembles. This specified physical pion mass was then used to identify the quark mass $m_e^*$ for each quenched ensemble, corresponding to the physical pion mass for that ensemble, via quadratic interpolation. For the given quenched ensembles, we defined the mass point by the physical pion mass of:
\begin{equation}
    M_\pi^* \equiv 780\,\text{MeV}.
\end{equation}
The resulting quark masses $m_e^*$ allowed us to quadratically interpolate the pion and vector correlation functions for each ensemble to the same mass point. To account for systematic errors in the quadratic interpolation, we performed an additional linear interpolation to $M_\pi^*$ using the data of the two nearest pion masses for each ensemble. The systematic error on the correlators was then determined by calculating the square root of the quadratic difference between the results of the linear and quadratic interpolations. We emphasize that there is flexibility in choosing the quantity used to define the same physical mass point for all ensembles. Instead of the physical pion mass $M_\pi^*$, we could also have used dimensionless combinations such as $\sqrt{t_0} M_\pi^*$. \\
As a cross-check to test the sensibility of interpolating correlators as described above, we determined the physical mass of the interpolated pion correlators and verified whether it matched $M_\pi^* = 780\,\text{MeV}$. As shown in \cref{fig:pion_masses}, we were able to retrieve the original mass for all quenched ensembles within the margin of error. 

\subsection{Blinding procedure}
Before the final datasets were generated a blinding procedure was put in place.  Each author performed a separate analysis with a common blinding factor applied to the correlators.  The blinding factor was produced by a non-invertible hash function that was known to one of the authors.  The blinding factors themself were not known to the authors.  After cross-checks were performed between the two analyses in a blinded manner, the result was fully unblinded by evaluating the hash function in a Zoom meeting on July 22, 2024.

\subsection{Local- and conserved-current correlators}\label{sec:correlators}
\begin{figure}
    \centering
    \includegraphics[width=.8\textwidth]{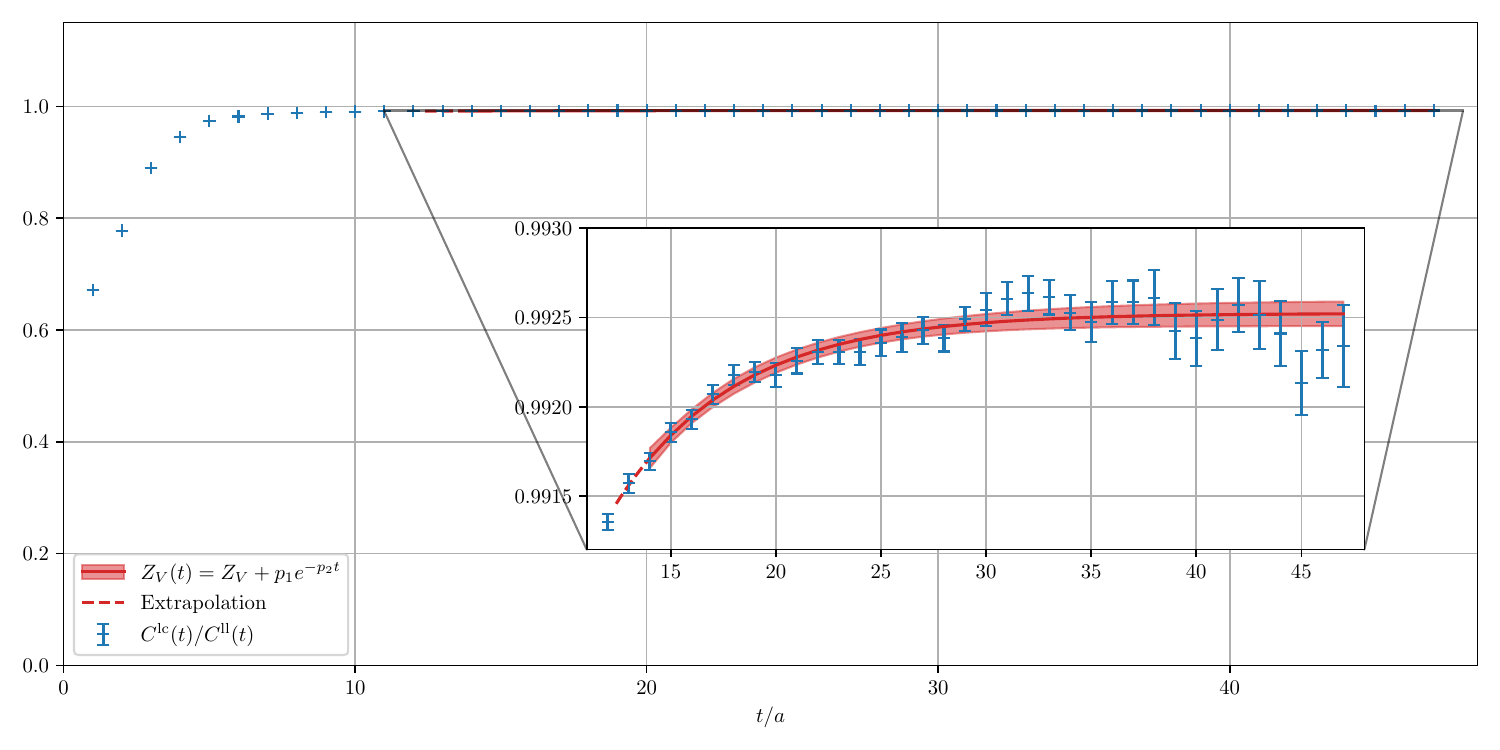}
    \caption{Plateau fit of the ratio $C^\text{lc}(t)/C^\text{ll}(t)$ using the model $Z_V(t) = Z_V + p_1 e^{-p_2 t}$. This is shown for the quenched $\beta=3.4$ ensemble with a lattice volume of $48^3 \times 196$.} 
    \label{fig:Z_V_fit_beta3p4}
\end{figure}

In our analysis, we consider the local lattice vector correlator  $J_\mu$ \cref{eq:VectorCurrent}, which we will refer to as  $J_\mu^l$  hereafter, as well as the conserved lattice vector current  $J_\mu^c$  as defined in \cite{blum2016domainwallqcd}. We study the correlators
\begin{align}
    C^{ab}(t) = \frac{1}{3} \sum_{\vec{x}} \sum_{j=0,1,2} \langle J_j^b(\vec{x},t) J_j^a(0)\rangle,
\end{align}
considering their local-local ($C^\text{ll}$) and local-conserved ($C^\text{lc}$) versions. For both the quenched and dynamical ensembles, we consider the up and down quark-connected contributions to the correlators in the isospin-symmetric limit. For the dynamical correlators an all-mode-averaging procedure \cite{DeGrand2005lowmodeaveraging,Bali2010noisereduction,Blum2013variancereduction,shintani2015covarianceapprox} is applied. In case of $C^\text{ll}$, we use a numerically cheaper but less precise estimator, $C^\text{ll}_\text{sloppy}$. We compute the difference $C^\text{ll} - C^\text{ll}_\text{sloppy}$ using a single source position for each sample element and then add this difference to our full-statistics estimator of $C^\text{ll}_\text{sloppy}$, which uses multiple sources, to obtain an estimate for $C^\text{ll}$. Similarly, we take advantage of the high correlations between the local-local and local-conserved correlators for a specific source position. To estimate $C^\text{lc}$, we first evaluate a sloppy estimator by calculating $C^\text{lc}_\text{sloppy} - C^\text{ll}_\text{sloppy}$ for a few correlated source positions and adding the full-statistics of the computationally cheaper estimator of $C^\text{ll}_\text{sloppy}$. Subsequently, the difference $C^\text{lc} - C^\text{lc}_\text{sloppy}$ is computed using a single source position for each sample element and added to the sloppy estimate. (The second analysis group used the ratio $C^\text{lc}/C^\text{ll}$ with smaller statistics combined with the full-statistics estimator of $C^\text{ll}$ to construct the $C^\text{lc}$ estimator.) \\
To determine the renormalization factor $Z_V$, we fit a plateau to the ratio $C^\text{lc}/C^\text{ll}$. For dynamical ensembles, we also compute an all-mode-average estimator for the latter quantity. The numerator $C^\text{lc}$ is estimated using a sloppy full-statistics estimator, which is then corrected by the difference between $C^\text{lc}$ and the sloppy estimator computed for a few point sources. We estimate the denominator $C^\text{ll}$ using the difference $C^\text{ll} - C^\text{ll}_\text{sloppy}$, with the full-statistics estimator $C^\text{ll}_\text{sloppy}$ restricted to the source positions from which the numerator $C^\text{lc}$ is computed. \\
For quenched ensembles with periodic BC, each correlator was estimated using three $Z_2$ wall sources at lattice time $0$. For quenched ensembles with open BC, the correlators were estimated using three such sources located at $T/2a$. The only exception is the ensemble with $\beta = 3.0$ and a lattice volume of $32^3 \times 96$, where the wall sources were placed at the beginning of the bulk. With the latter exception, the correlators for all ensembles were folded by keeping $t=0$ fixed and averaging the time slices $t$ and $T/a-t$ for $t = 1,...,T/2a-1$. 
For quenched periodic BC ensembles, we use a constant fit ansatz: 
\begin{align}
    Z_V(t) = Z_V
\end{align}
with fit parameter $Z_V$. The starting time slice of the fit range is chosen so that the extrapolated value of the previous time slice is within approximately $1\sigma$ distance to the actual data point at that time slice. The final time slice of the fit range is initially set to the last available time slice of the folded correlator. If the predicted value of the last time slice has more than $1\sigma$ tension with the actual data point, the final time slice is moved to earlier slices until the data point agrees with the prediction within the margin of error. For open BC quenched ensembles with a lattice volume of $24^3 \times 96$, we also consider a constant fit model. While the first time slice is chosen in the same way as for the periodic BC case, we set the initial final time slice to the end of the bulk. For quenched open BC ensembles with a lattice volume of $32^3 \times 96$ and $48^3 \times 192$, we use the fit ansatz
\begin{align}
    Z_V(t) = Z_V + p_1 e^{-p_2 t}
\end{align}
with fit parameters $Z_V$, $p_1$ and $p_2$. The fit range is chosen in the same way as for the other quenched open BC ensembles. An exemplary fit for $\beta=3.4$ is shown in \cref{fig:Z_V_fit_beta3p4}. To determine the normalization factors for the dynamical ensembles, we use a constant fit model. The starting time slice of the fit range is chosen using the same criterion as described above. However, due to increasing statistical noise at larger time slices, the final time slice is chosen such that the signal-to-noise ratio of the folded correlator remains larger than 10. \\
The vector correlators $C^{ll}$ and $C^{lc}$ are renormalized by multiplying them with a factor of $Z_V^2$ and $Z_V$, respectively. In the following discussions, we always consider the renormalized versions of the vector correlators.

\subsection{Short-distance window $\amuSD$}\label{sec:ShortDistanceWindow}
\begin{figure}
    \centering
    \includegraphics[width=\textwidth]{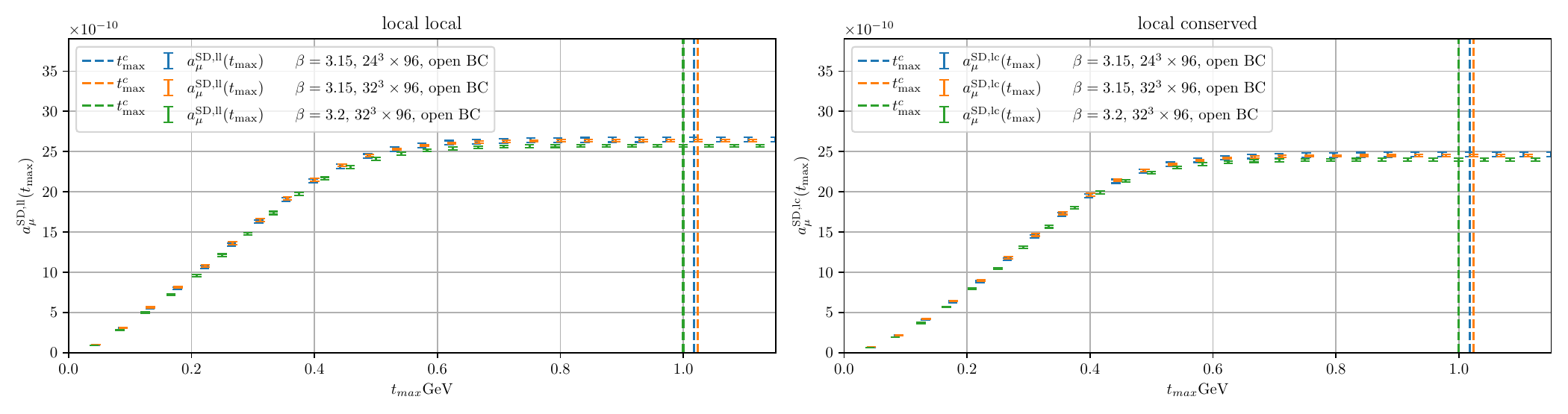}
    \caption{Exemplary representation of the blinded weighted sums $a_\mu^{\text{SD},i,\text{blinded}}(t_\text{max})$ computed with the weights $w_t$ for $\beta=3.15$ with lattice volume $24^3 \times 96$ and for $\beta=3.15$ and $\beta = 3.2$ with lattice volume $32^3 \times 96$. The chosen cutoffs $t_\text{max}^c$ are indicated with dashed lines. One the left, the blinded local-local measurements ($i=\text{ll}$) are shown. On the right, the blinded local-conserved measurements ($i = \text{lc}$) are displayed.}
    \label{fig:a_mu_sd_cutoff}
\end{figure}

To compute the short-distance window of the leading order HVP $\amuSD$, we need to evaluate the weighted sums:
\begin{align}
    a_\mu^{\text{SD},i} = \sum_{t=0}^\infty C^i(t) w_t^\text{SD},
\end{align}
where we introduced
\begin{align}
    w_t^\text{SD} = w_t [1 - \Theta(t, t_0=0.4\,\text{fm}, \Delta=0.15\,\text{fm})]
\end{align}
and $C^i(t)$ denotes the renormalized vector correlators with $i \in \lbrace\text{ll},\text{lc}\rbrace$. Additionally, we compute the same quantity with $w_t^\text{SD}$ replaced by $\hat{w}_t^\text{SD}$ which relies on the alternative weight definition from \cref{eq:w_hat}. The integrals in the weight computation, are evaluated numerically. For the physical muon mass and fine structure constant, we considered the values provided in \cite{particledatagroup2022pth}:
\begin{align}
m_\mu = 0.1056583755(23) \text{ GeV}, \quad \text{and} \quad \alpha = 0.0072973525693(11).
\end{align}
Due to the finite temporal extent and bulk sizes of our lattices, a cutoff $t_\text{max}$ is introduced in the weighted sum to study the resulting limitations:
\begin{align}
    a_\mu^{\text{SD},i}(t_\text{max}) = \sum_{t=0}^{t_\text{max}} C^i(t) w_t^\text{SD}.
\end{align}
For periodic BC ensembles, the maximum time slice included in the sum $t_\text{max}^c$ is set to half the temporal lattice extent. In case of ensembles with open BC, the maximum time slice is determined by the bulk region. An exception to this rule applies to the open BC ensembles with a lattice volume of $32^3 \times 96$. For $\beta = 3.0$ and $\beta = 3.15$, the cumulative sums are directly compared with their small volume counterparts (see \cref{fig:a_mu_sd_cutoff}). Since no boundary effects are observable when comparing the volumes, we set the maximum time slice for the large volumes to the same value as for the smaller volumes. In the case of $\beta = 3.2$, no small volume counterpart exists, so we choose $t_\text{max}^c$ such that the physical size of the bulk is the same or smaller as for the large volume at $\beta = 3.15$. To account for the systematics introduced by the cutoff choice, we include an additional systematic error, calculated as the absolute difference between $a_\mu^{\text{SD},i}(t_\text{max}^c)$ and its value evaluated 5 time slices earlier. 

\section{Continuum limit of the short-distance window $\amuSD$}\label{sec:ContinuumLimit}
The main goal of this work is to compute the continuum limit of the short-distance contribution $\amuSD$ to the leading-order HVP $a_\mu$. To achieve this, we rely on the expectation that discretization effects in both dynamical and quenched simulations are similar, and that the short-distance window is mostly insensitive to variations in quark masses \cite{blum2023update}. We propose performing a precise continuum limit using numerous quenched ensembles to get a quenched approximation of the short-distance contribution $\amuSDq$. The parameters determined from this study will then be used as a prior to constrain the parameters of a separate dynamical continuum extrapolation. The advantage of this approach is that the quenched ensembles necessary to obtain a precise estimate of $\amuSDq$ are computationally inexpensive compared to the corresponding dynamical estimates. The estimate of $\amuSD$ can be determined by employing only a small number of expensive dynamical ensembles, matched with their quenched counterparts in terms of scale.

\subsection{Quenched approximation}\label{sec:QuenchedApproximation}
\begin{figure}
    \centering
    \includegraphics[width=\textwidth]{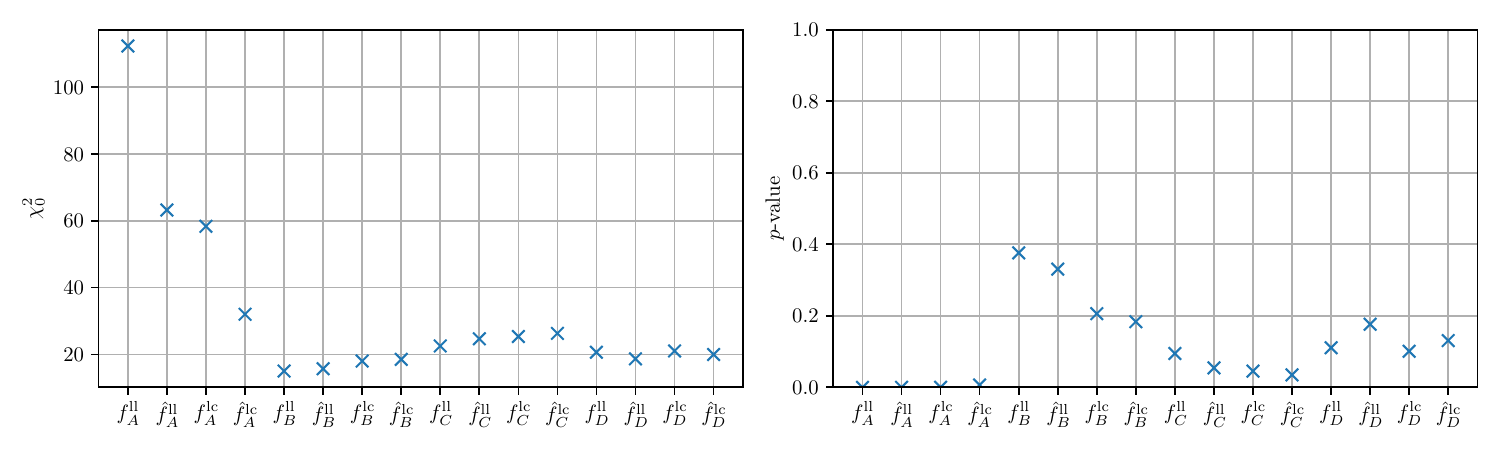}
    \caption{Overview of the computed $\chi^2$-function minima $\chi^2_0$ (left) and $p$-values (right) for $a_\mu^{\text{SD,q,blinded}}$. The fit models used for both weights $w_t$ and $\hat{w}_t$ are $f_A^i, f_B^i, f_C^i$, and $f_D^i$ as defined in \cref{eq:a_mu_sd_q_fit_model_fA,eq:a_mu_sd_q_fit_model_fB,eq:a_mu_sd_q_fit_model_fC,eq:a_mu_sd_q_fit_model_fD}. Fit results based on $\hat{w}_t$ are indicated by a hat.}
    \label{fig:a_mu_sd_q_fits_chi2_pval}
\end{figure}

\begin{figure}
    \centering
    \includegraphics[width=\textwidth]{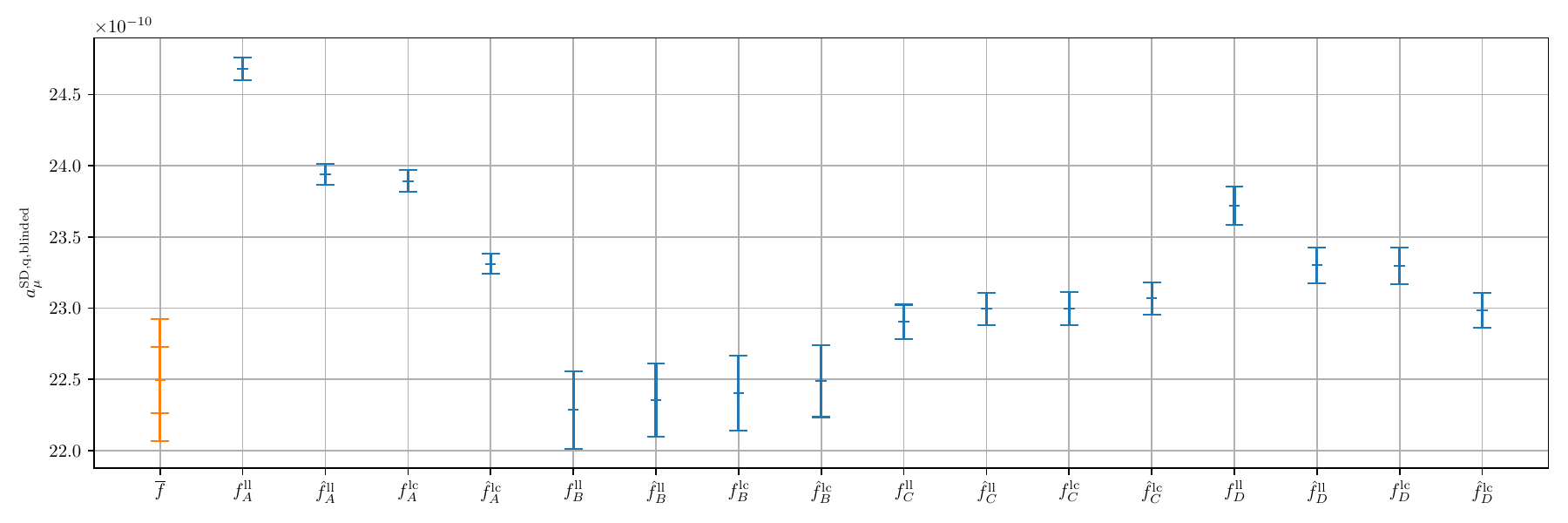}
    \caption{Comparison of the model average for $a_\mu^{\text{SD,q,blinded}}$, indicated by $\overline{f}$, with the estimates obtained from the quenched continuum extrapolations. The fit models used for both weights $w_t$ and $\hat{w}_t$ are $f_A^i, f_B^i, f_C^i$, and $f_D^i$ as defined in \cref{eq:a_mu_sd_q_fit_model_fA,eq:a_mu_sd_q_fit_model_fB,eq:a_mu_sd_q_fit_model_fC,eq:a_mu_sd_q_fit_model_fD}. Fit results based on $\hat{w}_t$ are indicated by a hat.}
    \label{fig:a_mu_sd_q_model_average}
\end{figure}

\begin{figure}
    \centering
    \includegraphics[width=\textwidth]{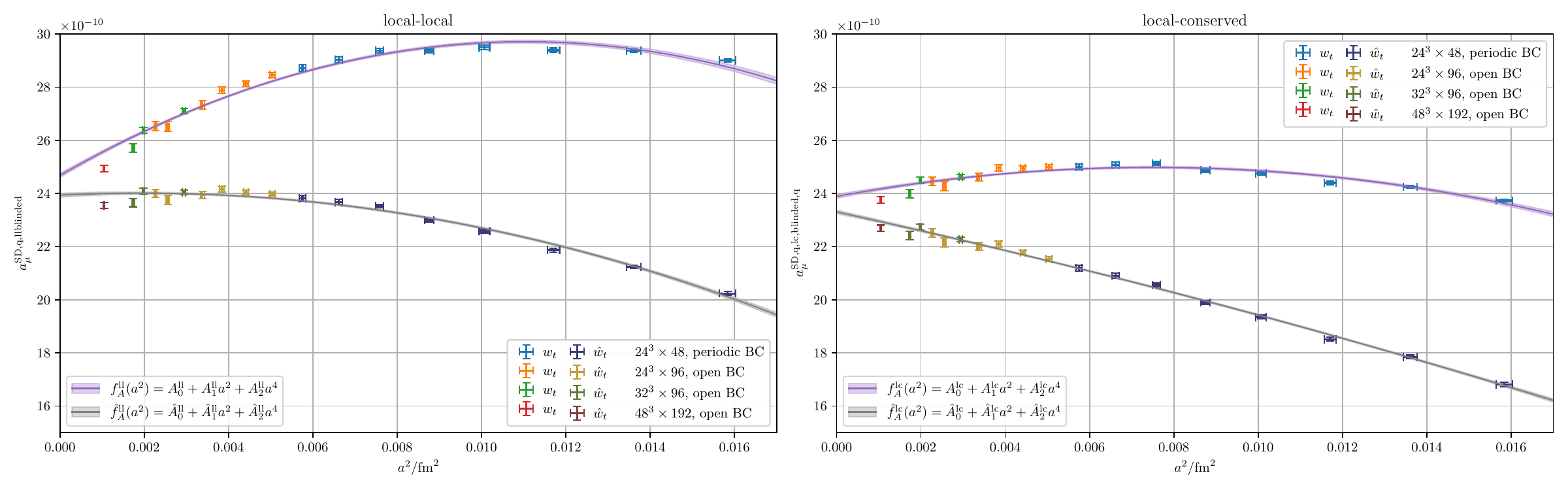}
    \caption{Continuum extrapolations of $\amuSDq$ computed with $w_t$ and $\hat{w}_t$ using the polynomial fit model $f_A^i(a^2)$ from \cref{eq:a_mu_sd_q_fit_model_fA}. The model fit based on the $\hat{w}_t$ is denoted by $\hat{f}_A^i(a^2)$. One the left, we show the fits of the blinded local-local measurements, i.e. $i=\text{ll}$. On the right, we display the fits of the blinded local-conserved measurements with $i=\text{lc}$.}
    \label{fig:a_mu_sd_q_fit_fA}
\end{figure}

\begin{figure}
    \centering
    \includegraphics[width=\textwidth]{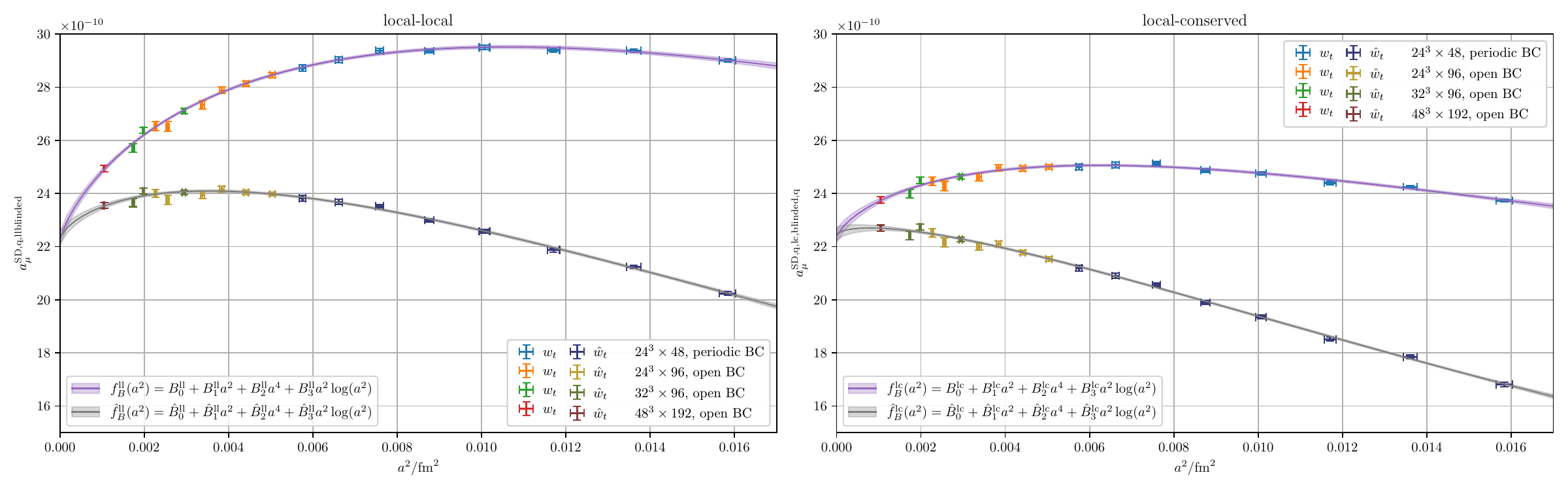}
    \caption{Continuum extrapolations of $\amuSDq$ computed with $w_t$ and $\hat{w}_t$ using the polynomial fit model $f_B^i(a^2)$ from \cref{eq:a_mu_sd_q_fit_model_fB} including a logarithmic term. The model fit based on the $\hat{w}_t$ is denoted by $\hat{f}_B^i(a^2)$. One the left, we show the fits of the blinded local-local measurements, i.e. $i=\text{ll}$. On the right, we display the fits of the blinded local-conserved measurements with $i=\text{lc}$.  We stress that the agreement in the continuum limit between the local-local and local-conserved extrapolations was not enforced in the fit but is rather used as a check of consistency.}
    \label{fig:a_mu_sd_q_fit_fB}
\end{figure}

\begin{figure}
    \centering
    \includegraphics[width=\textwidth]{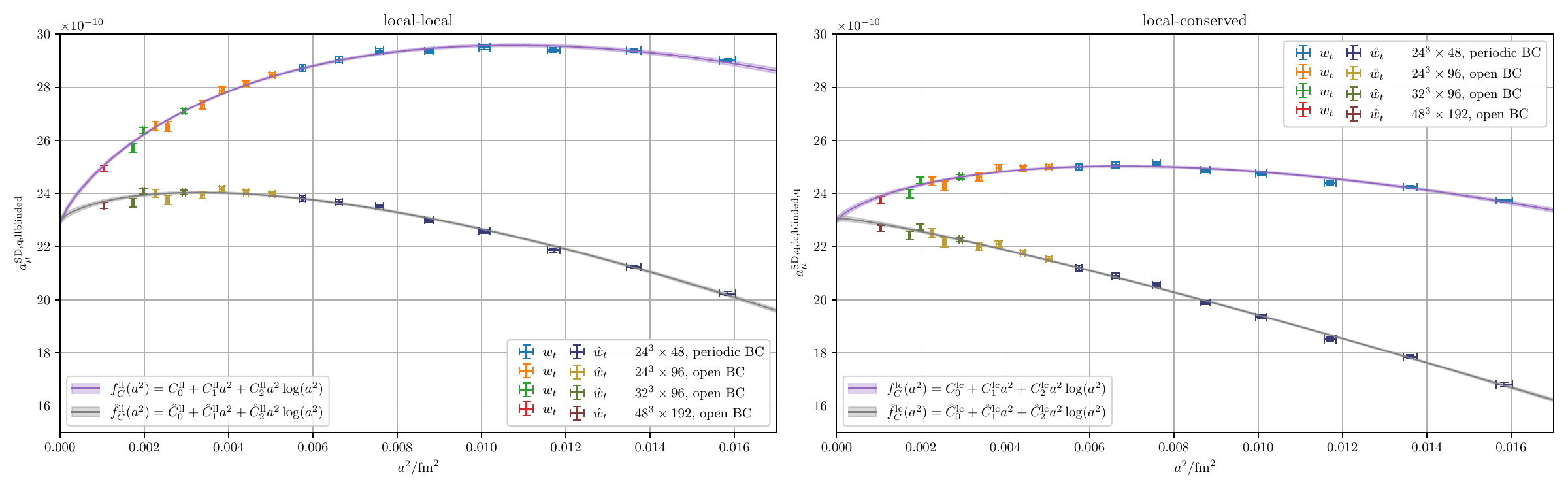}
    \caption{Continuum extrapolations of $\amuSDq$ computed with $w_t$ and $\hat{w}_t$ using the polynomial fit model $f_C^i(a^2)$ from \cref{eq:a_mu_sd_q_fit_model_fC} including a logarithmic term. The model fit based on the $\hat{w}_t$ is denoted by $\hat{f}_B^i(a^2)$. One the left, we show the fits of the blinded local-local measurements, i.e. $i=\text{ll}$. On the right, we display the fits of the blinded local-conserved measurements with $i=\text{lc}$.  We stress that the agreement in the continuum limit between the local-local and local-conserved extrapolations was not enforced in the fit but is rather used as a check of consistency.}
    \label{fig:a_mu_sd_q_fit_fC}
\end{figure}

\begin{figure}
    \centering
    \includegraphics[width=\textwidth]{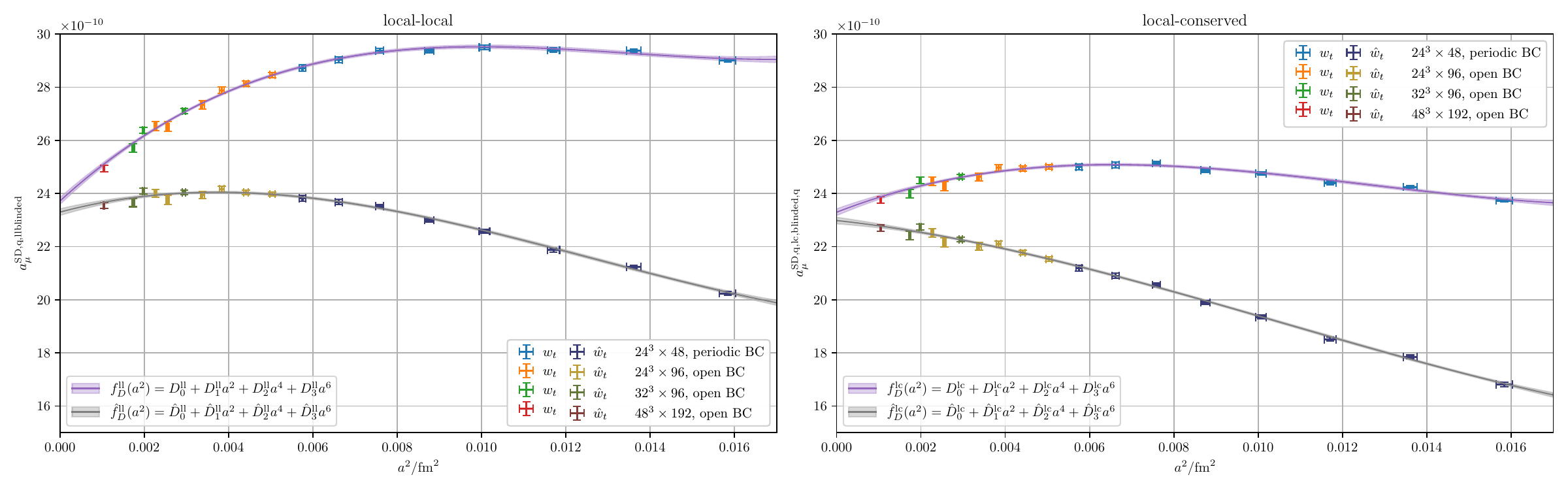}
    \caption{Continuum extrapolations of $\amuSDq$ computed with $w_t$ and $\hat{w}_t$ using the polynomial fit model $f_D^i(a^2)$ from \cref{eq:a_mu_sd_q_fit_model_fD}. The model fit based on the $\hat{w}_t$ is denoted by $\hat{f}_A^i(a^2)$. One the left, we show the fits of the blinded local-local measurements, i.e. $i=\text{ll}$. On the right, we display the fits of the blindaed local-conserved measurements with $i=\text{lc}$.}
    \label{fig:a_mu_sd_q_fit_fD}
\end{figure}

We start by discussing the continuum limit of the up and down quark-connected contributions to $\amuSDq$. The continuum extrapolations of $a_\mu^{\text{SD,q},i}$ are performed using separate fits for $i \in \lbrace \text{ll},\text{lc}\rbrace$ and for the weights $w_t$ and $\hat{w}_t$. Quantities determined with $\hat{w}_t$ are indicated with a hat. As described in \cref{sec:DataDescription}, there are 21 independent data points available for both local-local and local-conserved measurements, respectively. For the inverse couplings $\beta = 2.8, 3.0, 3.15$, we choose to only include the measurements corresponding to the larger lattice volume, reducing the number to 18. The following four fit ansaetze
\begin{subequations}
\begin{align}
    f_A^i(a^2) &= A_0^i + A_1^i a^2 + A_2^i a^4, \label{eq:a_mu_sd_q_fit_model_fA} \\
    f_B^i(a^2) &= B_0^i + B_1^i a^2 + B_2^i a^4 + B_3^i a^2 \log a^2, \label{eq:a_mu_sd_q_fit_model_fB} \\
    f_C^i(a^2) &= C_{0}^i + C_1^i a^2 + C_2^i a^2 \log a^2, \label{eq:a_mu_sd_q_fit_model_fC} \\
    f_D^i(a^2) &= D_{0}^i + D_1^i a^2 + D_2^i a^4 + D_3^i a^6, \label{eq:a_mu_sd_q_fit_model_fD}
\end{align}
\end{subequations}
are considered with fit parameters $A_\cdot^i, B_\cdot^i, C_\cdot^i, D_\cdot^i$ and $i \in \lbrace \text{ll}, \text{lc} \rbrace$. The models $f_A^i$ and $f_D^i$ represent pure polynomial fit ansaetze, while $f_B^i$ and $f_C^i$ incorporate a logarithmic term. An overview of all found $\chi^2$-function minima $\chi^2_0$ and the corresponding $p$-values is provided in \cref{fig:a_mu_sd_q_fits_chi2_pval}, with the continuum results displayed in \cref{fig:a_mu_sd_q_model_average}. Upon analyzing the blinded fit results, we observe a poor fit quality for $f_A$ as displayed in \cref{fig:a_mu_sd_q_fit_fA}. This is indicated by the corresponding $p$-values: $p_A^{\text{ll}}, p_A^{\text{lc}}, \hat{p}_A^{\text{ll}}$ and $\hat{p}_A^{\text{lc}}$ all of which evaluate roughly to zero. Furthermore, we find a tension between the blinded local-local and local-conserved predictions, as well as between the values computed with $w_t$ and those computed with $\hat{w}_t$. In contrast, the blinded fit results for the model $f_B^i$, shown in \cref{fig:a_mu_sd_q_fit_fB}, exhibit a significantly better quality, as can be quantified by higher $p$-values: $p_B^{\text{ll}} \approx 0.38$, $p_B^{\text{lc}} \approx 0.21$, $\hat{p}_B^{\text{ll}} \approx 0.33$ and $\hat{p}_B^{\text{lc}} \approx 0.18$. Additionally, all continuum results for this model ansatz agree within the error margins. The fit models $f_C^i$ and $f_D^i$ yield $p$-values that fall between those of models $f_A^i$ and $f_B^i$, with $f_D^i$ demonstrating better fit quality between the two. The corresponding fits are visualized in \cref{fig:a_mu_sd_q_fit_fC} for $f_C^i$ and in \cref{fig:a_mu_sd_q_fit_fD} for $f_D^i$. However, the continuum results from $f_D^i$ are not consistent, whereas those from $f_C^i$ are. These observations suggest that including a logarithmic term is necessary to obtain consistent results for both local-local and local-conserved estimates. This confirms the expectation of a large logarithmic dependence which without additional subtraction is even present at tree level \cite{Harris:2021azd,Sommer:2022wac,Husung:2022kvi}. \\
To obtain an estimate for $\amuSDqblinded$, we employ a model averaging procedure. For this, we assign the following probability to each model $M$,
\begin{align}
    P(M) = \frac{\exp(-\text{AIC}_M/2)}{\sum_{M'} \exp(-\text{AIC}_{M'}/2)} \,,
\end{align}
where $\text{AIC}_M$ denotes the value of the Akaike information criterion \cite{Akaike:1974} for model $M$ with
\begin{align}
    \text{AIC} = 2k + \chi^2_0
\end{align}
and $k$ representing the number of independent model parameters. This criterion rewards fit quality while penalizing a large number of parameters. Subsequently, we compute the model average for a fit parameter $p$ as
\begin{align}
    \overline{p} = \sum_M P(M) p_M \,,
\end{align}
where $p_M$ is the determined fit parameter for fit model $M$. The systematic variance for the model averaging is given by 
\begin{align}
    \sum_M P(M) (p_M - \overline{p})^2 \,.
\end{align}
Using this approach, our blinded result for the quenched approximation of the up and down quark-connected contribution to $\amuSD$ is thus given by
\begin{align}
a_\mu^{\text{SD,q,blinded}} = 22.50 (23) (36) \times 10^{-10}.
\label{eq:amuSDqblinded}
\end{align}
The value in the first parenthesis represents the statistical error, while the one in the second parenthesis denotes the systematic error, which is dominated by the model averaging. In \cref{fig:a_mu_sd_q_model_average}, the model average is compared to the estimates for $a_\mu^{\text{SD,q,blinded}}$ from the individual fits.

\subsection{Dynamical estimate}
\begin{figure}
    \centering
    \includegraphics[width=\textwidth]{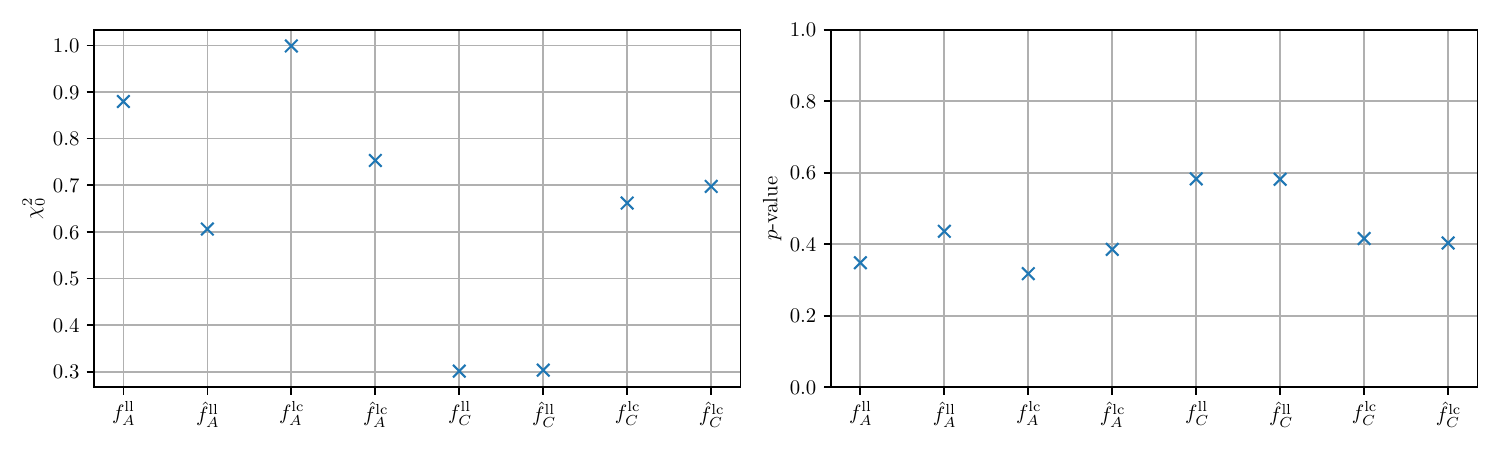}
    \caption{Overview of the computed $\chi^2$-function minima $\chi^2_0$ (left) and $p$-values (right) for $a_\mu^{\text{SD,blinded}}$. The fit models used for both weights $w_t$ and $\hat{w}_t$ are $f_A^i$ and $f_C^i$ as defined in \cref{eq:a_mu_sd_q_fit_model_fA,eq:a_mu_sd_q_fit_model_fC}. Fit results based on $\hat{w}_t$ are indicated by a hat.}
    \label{fig:a_mu_sd_dyn_fits_chi2_pval}
\end{figure}

\begin{figure}
    \centering
    \includegraphics[width=.9\textwidth]{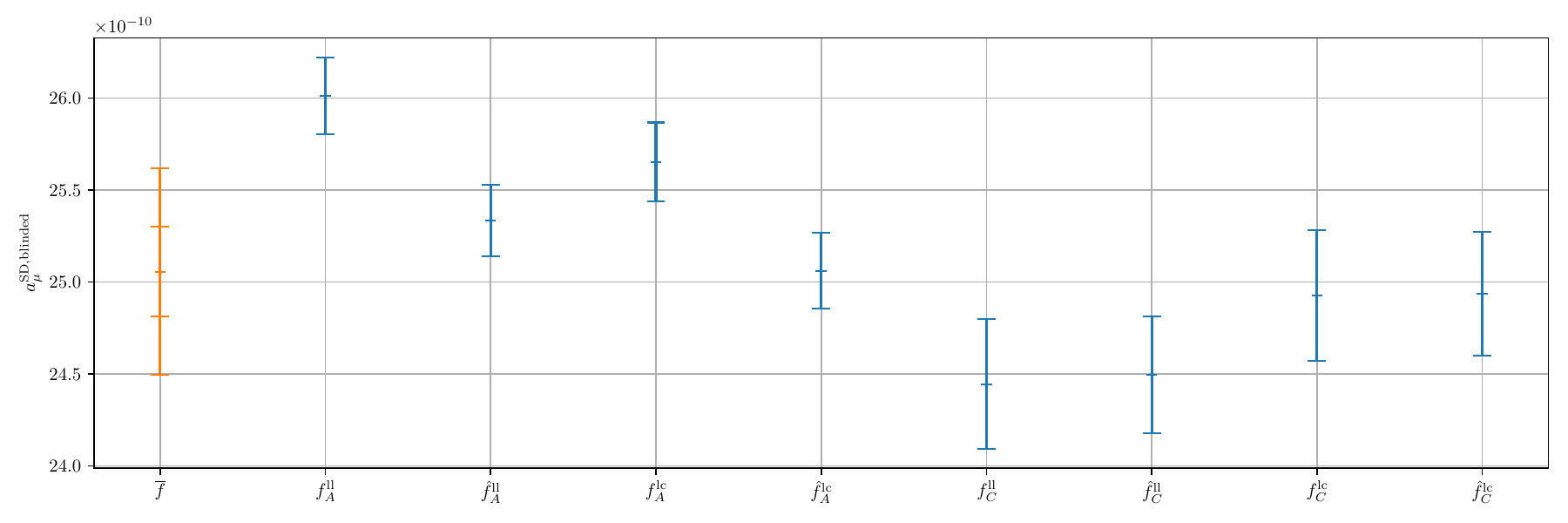}
    \caption{Comparison of the model average for $a_\mu^{\text{SD,blinded}}$, indicated by $\overline{f}$, with the estimates obtained from the continuum extrapolations. The fit models used for both weights $w_t$ and $\hat{w}_t$ are $f_A^i$ and $f_C^i$ as defined in \cref{eq:a_mu_sd_q_fit_model_fA,eq:a_mu_sd_q_fit_model_fC}. Fit results based on $\hat{w}_t$ are indicated by a hat.}
    \label{fig:a_mu_sd_dyn_model_average}
\end{figure}

\begin{figure}
    \centering
    \includegraphics[width=\textwidth]{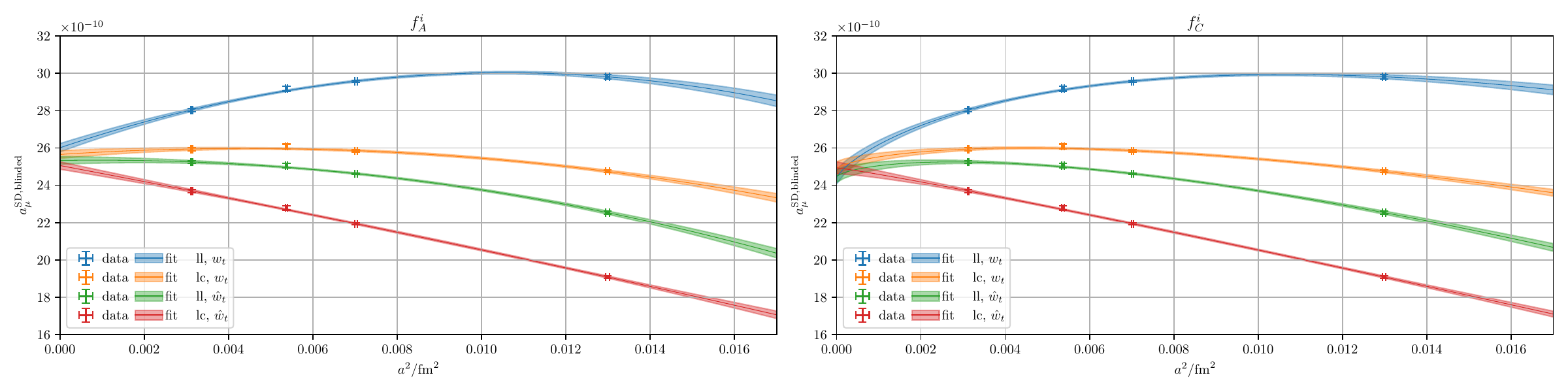}
    \caption{Continuum extrapolations of the blinded local-local (ll) and local-conserved (lc) versions of $\amuSD$ using the weights $w_t$ and $\hat{w}_t$. The left panel displays the fits for the model $f_A^i(a^2)$ as described in \cref{eq:a_mu_sd_q_fit_model_fA}, while the right panel shows the fits for the model $f_C^i(a^2)$ from \cref{eq:a_mu_sd_q_fit_model_fC}.}
    \label{fig:a_mu_sd_dyn_fits}
\end{figure}

\begin{figure}
    \centering
    \includegraphics[width=\textwidth]{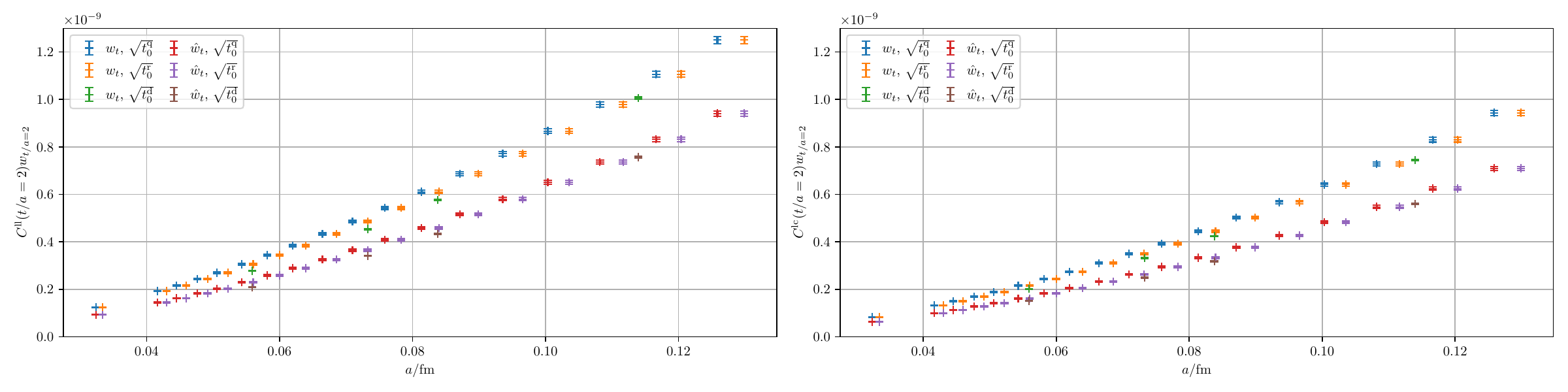}
    \caption{Comparison of the blinded $a_\mu^\text{SD}$ summand $C(t)w_t$ for $t/a=2$ between the dynamical ensembles with scale $\sqrt{t_0^\text{d}}$ and the quenched ensembles with the quenched scale $\sqrt{t_0^\text{q}}$ and the rescaled quenched scale $\sqrt{t_0^\text{r}}$. The summand is plotted versus the lattice spacing $a$ in fm for both weight definitions $w_t$ and $\hat{w}_t$. The left panel shows the local-local version, while the right panel displays the local-conserved version.}
    \label{fig:Ct_wt_t2}
\end{figure}

\begin{figure}
    \centering
    \includegraphics[width=\textwidth]{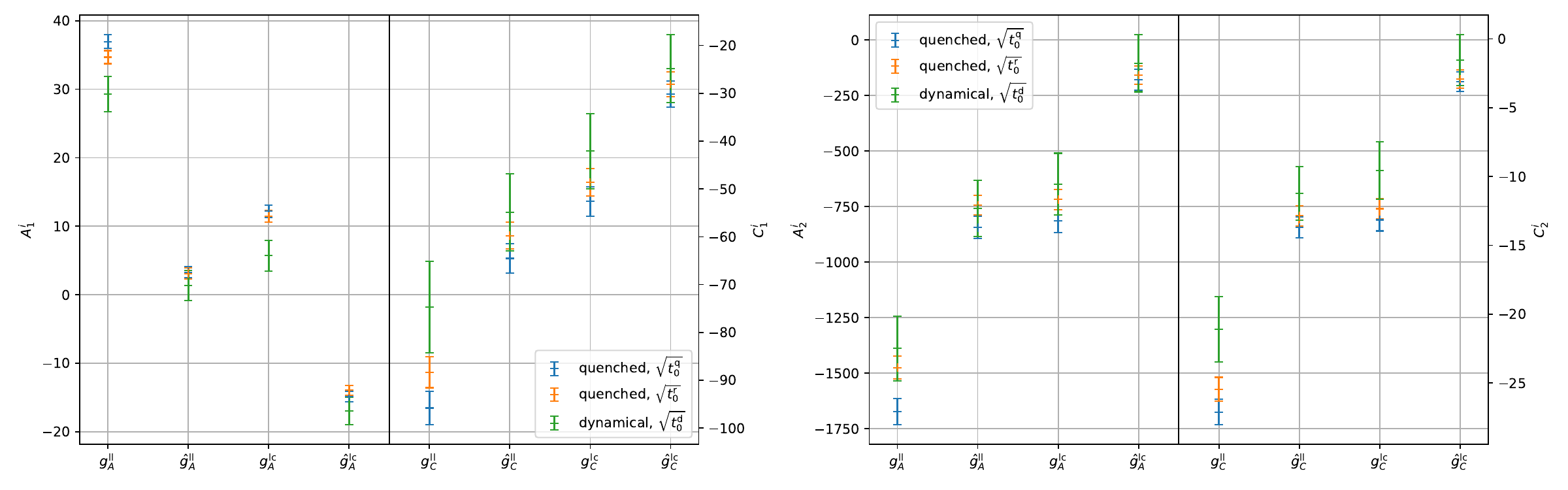}
    \caption{Comparison of the fit parameters for the models $g_A^i(a^2)$ and $g_C^i(a^2)$ applied to $\amuSDq$ using scales $\sqrt{t_0^\text{q}}$ (blue) and $\sqrt{t_0^\text{r}}$ (orange), and to $\amuSD$ using scale $\sqrt{t_0^\text{d}}$ (green). The left panel displays the first parameters $A_1^i$ and $C_1^i$ for both models, while the right panel shows the second parameters $A_2^i$ and $C_2^i$. Fits based on data computed with $\hat{w}_t$ are indicated by a hat.}
    \label{fig:compare_params}
\end{figure}

\begin{figure}
    \centering
    \includegraphics[width=\textwidth]{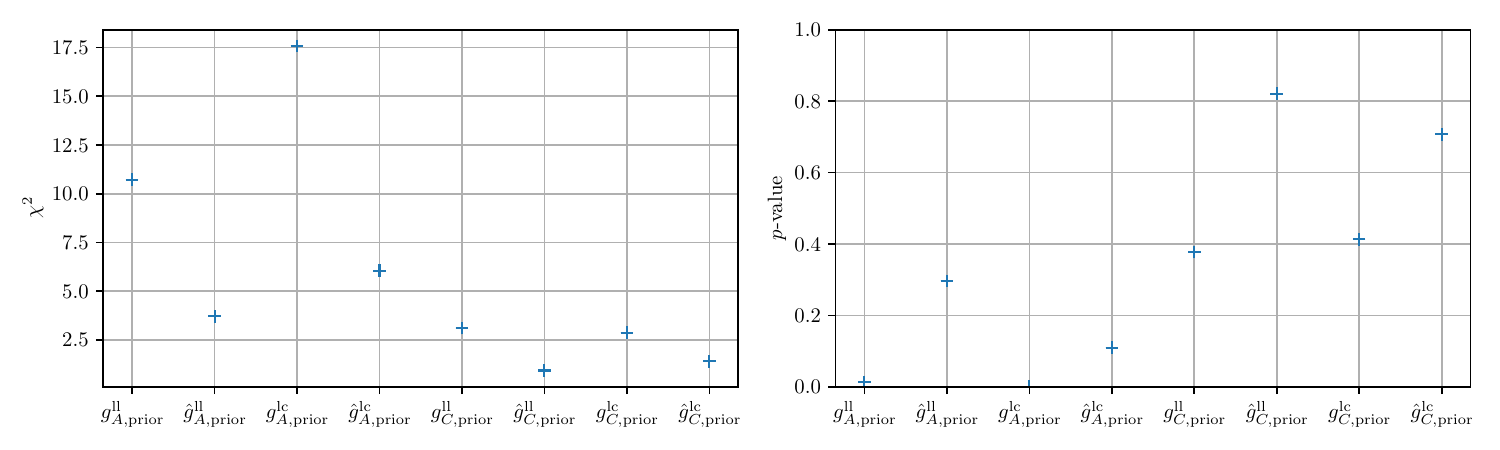}
    \caption{Overview of the computed $\chi^2$-function minima $\chi^2_0$ (left) and $p$-values (right) for $a_\mu^{\text{SD,blinded}}$. The fit models used for both weights $w_t$ and $\hat{w}_t$ are $g_A^i$ and $g_C^i$, as defined in \cref{eq:a_mu_sd_q_fit_model_gA,eq:a_mu_sd_q_fit_model_gC}, with additional constraints on the second and third parameters using priors. We denote the corresponding fits as $g_{\cdot,\text{prior}}^i$. Fit results based on $\hat{w}_t$ are indicated by a hat.}
    \label{fig:a_mu_sd_dyn_w_prior_fits_chi2_pval}
\end{figure}

\begin{figure}
    \centering
    \includegraphics[width=.9\textwidth]{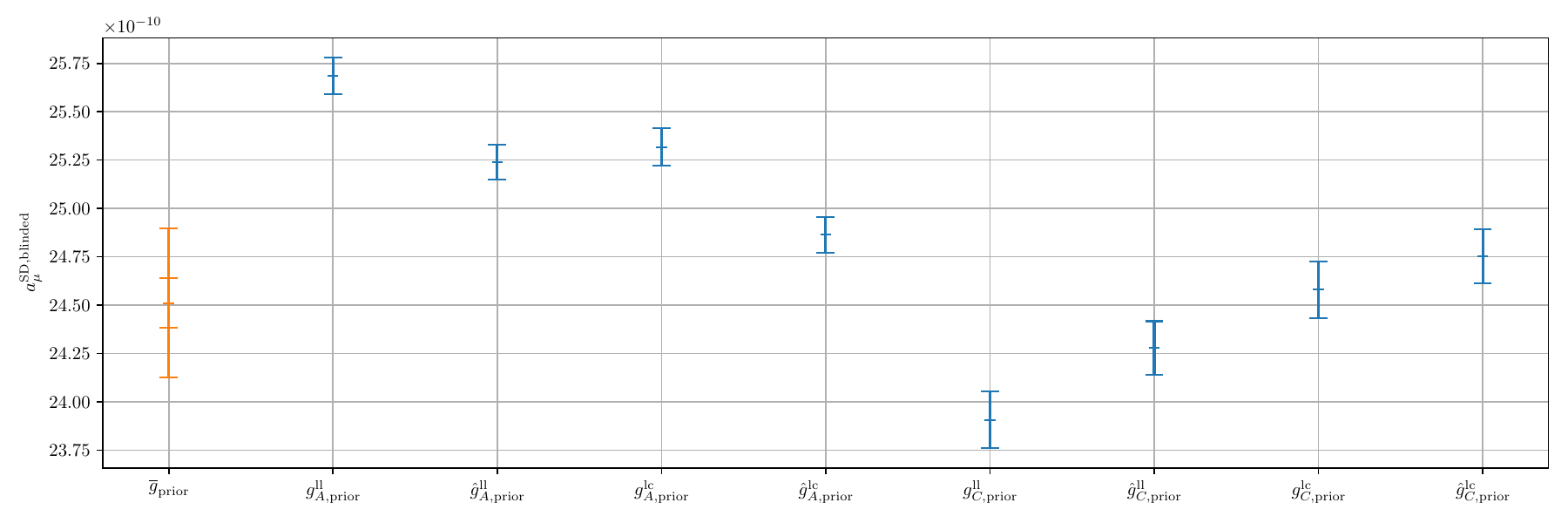}
    \caption{Comparison of the model average for $a_\mu^{\text{SD,blinded}}$, denoted as $\overline{g}_\text{prior}$, with the estimates obtained from the constrained continuum extrapolations. The fit models used for both weights $w_t$ and $\hat{w}_t$ are $g_A^i$ and $g_C^i$, as defined in \cref{eq:a_mu_sd_q_fit_model_gA,eq:a_mu_sd_q_fit_model_gC}, with additional constraints on the second and third parameters using priors. We denote the corresponding fits as $g_{\cdot,\text{prior}}^i$. Fit results based on $\hat{w}_t$ are indicated by a hat.}
    \label{fig:a_mu_sd_dyn_w_prior_model_average}
\end{figure}

\begin{figure}
    \centering
    \includegraphics[width=\textwidth]{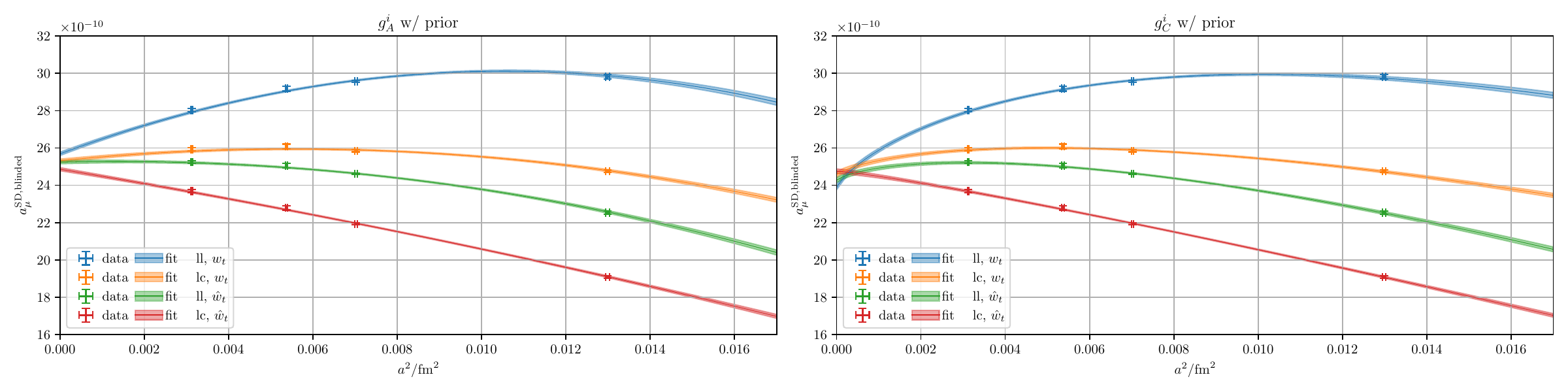}
    \caption{Constrained continuum extrapolations of the blinded local-local (ll) and local-conserved (lc) versions of $\amuSD$ using the weights $w_t$ and $\hat{w}_t$. The left panel displays the fits for the model $g_A^i(a^2)$ (see \cref{eq:a_mu_sd_q_fit_model_gA}) with priors, while the right panel shows the fits for the model $g_C^i(a^2)$ (see \cref{eq:a_mu_sd_q_fit_model_gC}) with priors. In both fits, the second and third parameters are constrained to the values of their quenched counterpart fits.}
    \label{fig:a_mu_sd_dyn_w_prior_fits}
\end{figure}

The difference in discretization errors between the quenched and dynamical short-distance contribution $\amuSD$ to the anomalous magnetic moment is suppressed by two-loop perturbative QCD effects. Since we want to use the quenched results as input for the dynamical continuum limit, the scale of the quenched simulations is matched with the dynamical scale, which includes sea quark effects. We optimize the matching by aligning the second time slice of the vector current, which effectively ensures the matching of the other short-distance time slices as well. This is achieved by rescaling the quenched scale $\sqrt{t_0^\text{q}}$ to match the dynamical scale $\sqrt{t_0^\text{d}}$. We refer to the rescaled quenched scale as $\sqrt{t_0^\text{r}}$. 
A comparison of the $\amuSD$ summand $C(t)w_t$ with $t=2$ for the three scales is displayed in \cref{fig:Ct_wt_t2}. This comparison indicates that as the lattice spacing approaches the continuum limit, the primary summands contributing to the short-distance contribution are comparable in scale. Consequently, the model parameters for both the dynamical and quenched continuum extrapolations are expected to be similar. Considering that only four dynamical ensembles (see \cref{tab:DynamicalData}) are included in this study, we are restricted to fitting models with three parameters: $f_A$ and $f_B$. Both models produce reasonable $p$-values ranging from $0.3$ to $0.6$, though the parameter variances are approximately $10\%$ and the continuum results exhibit a spread. The continuum extrapolations for models $f_A$ and $f_B$ are presented in \cref{fig:a_mu_sd_dyn_fits}, while \cref{fig:a_mu_sd_dyn_fits_chi2_pval} provides an overview of the $\chi^2$-function minima $\chi^2_0$ and $p$-values. The continuum values and the AIC model average $\overline{f}$ are illustrated in \cref{fig:a_mu_sd_dyn_model_average}. \\ 
When comparing the second and the third model parameters for the fits using the quenched scale $\sqrt{t_0^\text{q}}$, the rescaled quenched scale $\sqrt{t_0^\text{r}}$, and the dynamical scale $\sqrt{t_0^\text{d}}$, we found that these parameters agreed best when using the multiplicative fit ansatze: 
\begin{subequations}
\begin{align}
    g_A^i(a^2) &= A_0^i (1 + A_1^i a^2 + A_2^i a^4), \label{eq:a_mu_sd_q_fit_model_gA} \\
    g_C^i(a^2) &= C_{0}^i (1 + C_1^i a^2 + C_2^i a^2 \log a^2), \label{eq:a_mu_sd_q_fit_model_gC} \\
\end{align}
\end{subequations}
instead of \cref{eq:a_mu_sd_q_fit_model_fA,eq:a_mu_sd_q_fit_model_fC}. We remark that using $g_\cdot^i$ instead of $f_\cdot^i$ amounts to a mere reparametrization of the same discretization and yields the same continuum result. \cref{fig:compare_params} shows that the quenched fit parameters for the rescaled scale $\sqrt{t_0^\text{r}}$ are in good agreement with parameters for the dynamical scale $\sqrt{t_0^\text{d}}$ in almost all cases. In cases where the match is not as precise, the parameters corresponding to the rescaled scale are still more closely aligned with the dynamical model parameters than with the original quenched scale parameters. This insight motivates us to constrain the parameters of the dynamical continuum extrapolation with the parameters of the quenched fits by adding the priors
\begin{align}
    \sum_{k=1}^2 \frac{(A_k^i - A_k^{i,\text{q}})^2}{\text{var}(A_k^{i,\text{q}})} \quad \text{and} \quad \sum_{k=1}^2 \frac{(C_k^i - C_k^{i,\text{q}})^2}{\text{var}(C_k^{i,\text{q}})}
\end{align} 
to the respective $\chi^2$ functionals, where $A_k^{i,\text{q}}$ and $C_k^{i,\text{q}}$ denote the parameters of the corresponding quenched simulations with scale $\sqrt{t_0^\text{r}}$ and fit models $g_A^i$ and $g_C^i$. An overview of the corresponding $\chi^2$-function minima $\chi^2_0$ and $p$-values is provided in \cref{fig:a_mu_sd_dyn_w_prior_fits_chi2_pval}. The errors on the constrained continuum results are reduced to approximately the same order of magnitude as in the quenched case, as expected, while still yielding $p$-values greater than $0.1$ for all but two fits: $g_A^\text{ll}$ and $g_A^\text{lc}$. This outcome may be explained for these two cases by the above observed discrepancy between the fit parameters for the dynamical scale and the rescaled quenched scale, as shown in \cref{fig:compare_params}. The constrained continuum extrapolations are shown in \cref{fig:a_mu_sd_dyn_w_prior_fits}. The visible spread between the continuum results for the different discretizations and fit ansaetze is used in the AIC model averaging as an error estimate. Additionally, we note that, we also performed constrained dynamical fits using the multiplicative model versions of $f_B$ and $f_D$, defined as:
\begin{subequations}
\begin{align}
g_B^i(a^2) &= B_0^i (1 + B_1^i a^2 + B_2^i a^4 + B_3^i a^2 \log a^2), \label{eq:a_mu_sd_q_fit_model_gB} \\
g_D^i(a^2) &= D_{0}^i (1 + D_1^i a^2 + D_2^i a^4 + D_3^i a^6). \label{eq:a_mu_sd_q_fit_model_gD}
\end{align}
\end{subequations}
However, these models yielded vanishing $p$-values in our analysis and therefore do not contribute significantly to the model average. Consequently, our model average for the constrained dynamical fits comprises eight results, all of which are displayed in \cref{fig:a_mu_sd_dyn_w_prior_model_average}.

After all checks were completed, we unblinded the results on July 22.
Finally, we need a correction from the $\approx 280$ MeV pion mass of the dynamical ensembles used in this work to the physical pion mass, which were calculated for the RBC/UKQCD23 SD window in Ref.~\cite{blum2023update},
\begin{align}
  \Delta a_\mu^{\rm SD,~mass} = 0.41(4) \times 10^{-10} .
\end{align}
Our final unblinded result for the light-quark connected short-distance window in the isospin symmetric limit is
\begin{align}
a_\mu^{\rm ud,~iso,~SD} &= 47.62(0.32)_{\rm stat.}(0.60)_{\rm syst.} \times 10^{-10} \,.
\end{align}


\section{Conclusions and outlook}
\begin{figure}
    \centering
    \includegraphics[width=0.4\linewidth]{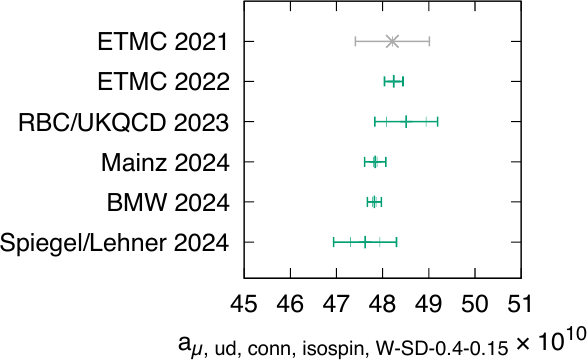}
    \caption{Comparison of our result with existing literature: ETMC 2021 \cite{Giusti:2021dvd}, ETMC 2022 \cite{ExtendedTwistedMass:2022jpw}, RBC/UKQCD 2023 \cite{RBC:2023pvn}, Mainz 2024 \cite{Kuberski:2024bcj}, BMW 2024 \cite{Boccaletti:2024guq}.} 
    \label{fig:comparison}
\end{figure}

In this study, we computed the short-distance Euclidean window of the hadronic vacuum polarization. To prevent potential bias towards previously published results, we employed a blinding procedure involving two independent analysis groups. The focus of this work was on the dominant quark-connected, light-quark, isospin-symmetric contribution, scrutinizing its continuum limit from first-principles Lattice QCD without perturbative input. A precise quenched study revealed the necessity of a logarithmic dependency to obtain consistent continuum results across different discretizations in agreement with the expectation \cite{Harris:2021azd,Sommer:2022wac,Husung:2022kvi}. We subsequently demonstrated that these quenched parameter results could be used to constrain the parameters of the corresponding dynamical continuum extrapolations. Our result for the short-distance window $a_\mu^{\rm ud,~iso,~SD}$, derived from the model average of these constrained continuum limits, is consistent with recently published results that rely on perturbative input, as shown in \cref{fig:comparison}. However, our error estimate is significantly larger compared to those results. \\ 
There are many directions one may explore in future work. For instance, one may include an additional dynamical ensemble with a finer lattice spacing compared to the currently available ones. This would allow us to achieve higher precision and further scrutinize the continuum limit. Additionally, one may systematically compare our results against perturbative calculations. This would involve tuning the lower cutoff of the short-distance window to ensure a robust and detailed comparison.

{\bf Acknowledgments.}  We thank our colleages from the RBC/UKQCD collaborations for valuable discussions.  We acknowledge computing resources provided at the University of Regensburg on the QPace3 and QPace4 clusters.  The authors gratefully acknowledge the Gauss Centre for Supercomputing
e.V. (www.gauss-centre.eu) for funding this project by providing
computing time on the GCS Supercomputer JUWELS at Jülich
Supercomputing Centre (JSC).   We
gratefully acknowledge disk and tape storage provided by the
University of Regensburg with support from the DFG.

\clearpage
\bibliography{references}

\end{document}